\pdfoutput=1     
\documentclass[12pt]{JINST}
\usepackage{graphicx}
\usepackage{amsmath,graphicx,epsfig,amssymb,dsfont,mathtools}
\usepackage{inputenc}
\usepackage{ulem} 
\usepackage{bigstrut}
\usepackage{makecell}
\usepackage{slashed}
\usepackage{multirow}
\usepackage{subfigure}
\usepackage{url}
\usepackage{diagbox}
\usepackage{etoolbox}
\usepackage{lineno}
\usepackage{diagbox}
\usepackage{rotating}
\usepackage{lscape}
\usepackage{array}

\title{Particle Identification Using Boosted Decision Trees in the Semi-Digital Hadronic Calorimeter Prototype}

\author{\centering 
\LARGE\bf The CALICE Collaboration}

\author{\centering
D.\,Boumediene,
\\ \it
Universit\'e Clermont Auvergne, Universit\'e Blaise Pascal, CNRS/IN2P3, LPC, 4 Av. Blaise Pascal, TSA/CS 60026,
F-63178 Aubi\`ere, France
}

\author{\centering
A.\,Pingault, 
M.\,Tytgat
\\ \it
Ghent University, Department of Physics and Astronomy,
Proeftuinstraat 86, B-9000 Gent, Belgium
}

\author{\centering 
B.\,Bilki, 
D.\,Northacker,
Y.\,Onel
\\ \it
University of Iowa, Dept. of Physics and Astronomy,
203 Van Allen Hall, Iowa City, IA 52242-1479, USA
}

\author{\centering 
G.\,Cho, 
D-W.\,Kim,
S.\,C.\,Lee, 
W.\,Park, 
S.\,Vallecorsa
\\ \it
Gangneung-Wonju National University
Gangneung 25457, South Korea
}

\author{\centering 
Y.\,Deguchi,
K.\,Kawagoe,
Y.\,Miura,
R.\,Mori,
I.\,Sekiya,
T.\,Suehara,
T.\,Yoshioka
\\ \it
Department of Physics and Research Center for Advanced Particle Physics,
Kyushu University, 744 Motooka, Nishi-ku, Fukuoka 819-0395, Japan
}

\author{\centering 
L.\,Caponetto, 
C.\,Combaret, 
R.\,Ete\footnote{Now at DESY}, 
G.\,Garillot, 
G.\,Grenier, 
J-C.\,Ianigro, 
T. \,Kurca,  
I.\,Laktineh, 
B.\,Liu$^{a}$ ,
B.\,Li,
N.\,Lumb, 
H.\,Mathez, 
L.\,Mirabito, 
A.\,Steen\footnote{Now at NTU}
\\ \it
Univ Lyon, Univ CLaude Bernard Lyon 1, 
CNRS/IN2P3, IP2I Lyon, F-69622 
Villeurbanne, France
}

\author{\centering 
E.\,Calvo Alamillo,
M.C.\,Fouz,
J.\,Marin,
J.\,Navarrete,
J.\,Puerta Pelayo,
A.\,Verdugo
\\ \it
CIEMAT, Centro de Investigaciones Energeticas, Medioambientales y Tecnologicas, Madrid, Spain 
}

\author{\centering 
F.\,Corriveau,
B.\,Freund\footnote{Also at Argonne National Laboratory}
\\ \it
Department of Physics, McGill University,
Ernest Rutherford Physics Bldg.,
3600 University Ave.,
Montr\'{e}al, Qu\'{e}bec,
Canada H3A 2T8
}

\author{\centering M.\,Chadeeva, M.\,Danilov\footnote{Also at MIPT} \\ \it P.\,N.\, Lebedev Physical Institute of the Russian Academy of Sciences, 53 Leninsky prospekt, Moscow, 119991 Russia }

\author{\centering 
L.\,Emberger,
C.\,Graf,
F.\,Simon,
C.\,Winter
\\ \it
Max-Planck-Institut f\"ur Physik,
F\"ohringer Ring 6,
D-80805 Munich, Germany
}

\author{\centering 
J.\,Bonis, 
D.\,Breton, 
P.\,Cornebise, 
A.\,Gallas, 
J.\,Jeglot, 
A.\,Irles, 
J.\,Maalmi, 
R.\,P\"oschl, 
A.\,Thiebault, 
F.\,Richard, 
D.\,Zerwas 
\\ \it
UniversitŽ Paris-Saclay, CNRS/IN2P3, IJCLab, 91405 Orsay, France 
}

\author{\centering 
M.\,Anduze,
V.\,Balagura,
V.\,Boudry,
J-C.\,Brient, 
E.\,Edy,
F.\,Gastaldi, 
R.\,Guillaumat,
F.\,Magniette,
J.\,Nanni,   
H.\,Videau
\\ \it
 Laboratoire Leprince-Ringuet (LLR)  -- CNRS, \'{E}cole polytechnique, Institut Polytechnique de Paris
 Palaiseau, F-91128 France
}

\author{\centering 
S.\,Callier,
F.\,Dulucq, 
Ch.\,de la Taille, 
G.\,Martin-Chassard,
L.\,Raux, 
N.\,Seguin-Moreau
\\ \it
Laboratoire OMEGA -- \'{E}cole Polytechnique-CNRS/IN2P3, 
Palaiseau, F-91128 France
}

\author{\centering 
J.\,Cvach, 
M.\,Janata, 
M.\,Kovalcuk, 
J.\,Kvasnicka, 
I.\,Polak, 
J.\,Smolik, 
V.\,Vrba, 
J.\,Zalesak, 
J.\,Zuklin
\\ \it
Institute of Physics, The Czech Academy of Sciences,
Na Slovance 2, CZ-18221 Prague 8, Czech Republic
}

\author{\centering
Y.Y. \,Duan,
 S.  \,Li, 
 J.  \,Guo,
  J.F. \, Hu, 
  F.  \,Lagarde,
  B.\,Liu$^{a}$, 
  Q.P. \, Shen, 
  X.  \,Wang, 
  W.H. Wu, 
  H.J.  \,Yang, 
  Y.F.  \,Zhu
\\ \it 
Tsung-Dao Lee Institute, Institute of Nuclear and Particle Physics, School of Physics and Astronomy, Shanghai Jiao Tong University,
Key Laboratory for Particle Physics, Astrophysics and Cosmology (Ministry of Education),
Shanghai Key Laboratory for Particle Physics and Cosmology,
800 Dongchuan Road, Shanghai, 200240, P. R. China
}

\author{
\it
$^{a}$ Corresponding author\newline
E-mail: \email{b.Liu@ipnl.in2p3.fr, 610412075@sjtu.edu.cn}
}

\abstract{The CALICE Semi-Digital Hadronic CALorimeter (SDHCAL) prototype using Glass Resistive Plate Chambers as a sensitive medium is the first technological prototype of a family of high-granularity calorimeters developed by the CALICE collaboration to equip the experiments of future leptonic colliders.  It was exposed to beams of hadrons, electrons and muons several times in the CERN PS and SPS beamlines  between 2012 and 2018. We present here a new method of  particle identification within the SDHCAL using the Boosted Decision Trees (BDT) method applied to the data collected in 2015.  The performance of the method is tested first with Geant4-based simulated events and then on the data collected by the SDHCAL in the energy range between 10 and 80~GeV with 10~GeV energy steps. The BDT method is then used to reject  the electrons and muons that contaminate the SPS hadron beams.}


\begin{document}
\maketitle
\flushbottom
\linenumbers

\label{sec:intro}
\section{Introduction}

The Semi-Digital Hadronic CALorimeter (SDHCAL)~\cite{SDHCAL} is the first of a series of technological high-granularity prototypes developed by the CALICE collaboration. These technological prototypes  have  their  readout electronics embedded in the detector and they are power-pulsed to reduce the power consumption in experiments proposed within the International Linear Collider (ILC) project~\cite{ILC}. The mechanical structure of these prototypes is part of their absorber. All these aspects increase the compactness of the calorimeters and improve their suitability to apply the Particle Flow Algorithm (PFA) techniques~\cite{PFA1, PFA2, PFA3}. The SDHCAL is made of 48 active layers, each of them equipped with a 1~m~$\times$~1~m Glass Resistive Plate Chamber (GRPC) and  an Active Sensor Unit (ASU) of the same size hosting on one face (the one in contact with the GRPC)~pickup pads of 1~cm~$\times$~1~cm and 144 HARDROC2 ASICs~\cite{HARD} on the the other face. The GRPC and the ASU are assembled within a cassette made of two stainless steel plates, 2.5~mm thick each. The 48 cassettes are inserted in a self-supporting mechanical structure made of 51 plates, 15~mm thick each, of the same material as the cassettes, bringing the total absorber thickness to 20~mm per layer. The empty space between two consecutive plates is 13~mm to allow the insertion of one cassette of 11~mm thickness. The HARDROC2 ASIC has 64 channels to read out 64 pickup pads. Each channel has three parallel digital circuits whose parameters can be configured to provide 2-bit encoded information indicating if the charge seen by each pad has passed any of the three different thresholds associated to each digital circuit. This multi-threshold readout is proposed to improve on the energy reconstruction of hadronic showers at high energy ($> 30$~GeV) with respect to the simple binary readout mode as explained in Ref.~\cite{FirstResults}.

The SDHCAL was exposed several times to different kinds of particle beams in the CERN PS and SPS beamlines between 2012 and 2018. The energy reconstruction of hadronic showers within the SDHCAL using the associated number of fired pads with multi-threshold readout information is presented in Ref.~\cite{FirstResults}. The contamination of the SPS hadron beams such as electrons and muons and the absence of Cherenkov counters during the data taking require the use of the event topology  to select the hadronic events before reconstructing their energy. Although the rejection of muons based on the average number of hits per crossed layer is efficient, the rejection of electrons is more difficult because some hadronic showers behave in similar way as the electromagnetic ones in particular at low energy.  To reject the electron events, the analysis presented in Ref.~\cite{FirstResults} requires the shower to start after the fifth layer. Almost all of the electrons are expected to start showering before crossing the equivalent of 6 radiation lengths ($X_0$)~\footnote{The longitudinal depth of the SDHCAL prototype layer is about 1.2~$X_0$.}.  Although this selection is found to have no impact on the hadronic energy reconstruction, it represents 0.6 interaction length ($\lambda_I$) and thus reduces the amount of the hadronic showers available for analysis.

In this paper we explore another method  to reject the electron and muon contaminations, that is not based on the shower start requirement and does thus preserve the statistics.
The new method is based on Boosted Decision Trees~(BDT)~\cite{BDT,BDT1}, a part of so-called  MutiVariate Analysis~(TMVA) technique~\cite{MVAT}. 
In the BDT, different variables associated to the topology of the event are exploited in order to distinguish between the hadronic and the electromagnetic showers, and also to identify muons including radiative ones that may exhibit a shower-like shape. In this paper, section~\ref{Para:sample} introduces the simulation and beam data samples which are used to study the performance of both the BDT and the standard method 
described in Ref.~\cite{FirstResults}. Section~\ref{Para:BDT} describes the selected input variables of BDT and the two approaches to build the classifier of BDT. Section~\ref{Para:results} presents the results of the hadron selection using BDT. Finally, section~\ref{Para:conclusion} gives the conclusion.

\section{Monte Carlo samples and beam data samples}
\label{Para:sample}
The SDHCAL prototype was exposed to pions, muons and electrons in the SPS of CERN in October 2015. In order to avoid GRPC saturation problems at high particle rate,  only runs with a particle rate smaller than 1000 particles/spill are selected for the analysis. In these conditions, pion events  at several energy points (10, 20, 30, 40, 50, 60, 70, 80~GeV) and muon events of 110~GeV were collected  as well as  electron events of  10, 15, 20, 25, 30, 40, 50~GeV.
While the electron and muon beams are rather pure, the pion beams are contaminated by two sources. One is the electron contamination despite the use of a lead filter to reduce the number of electrons. The other is the muon contamination resulting from pions decaying before reaching the prototype.
To apply the BDT method, six variables are selected and used in the Toolkit for MultiVariate  data Analysis (TMVA) package~\cite{MVAT} to build the decision tree.

To study the performance of the BDT method, we use  the Geant4.9.6 Toolkit package~\cite{GEANT4} associated to the FTF-BIC\footnote{The FTF model is based on the Fritiof description of string excitation and fragmentation. The BIC model uses Geant4 binary cascade for primary protons and neutrons with energies below~10~GeV. It describes the production of secondary particles produced in interactions of protons and neutrons with nuclei. }~\cite{RefFTF, RefBIC} physics list to generate pion, electron and muon events under the same conditions as in the beam test at CERN-SPS beamline. For the training of the BDT, 10k events for each energy point from 10~GeV to 80~GeV with a step of 10~GeV for pions, muons and electrons were produced.
The same amount of events of each species is produced and used to test the BDT method at the same time. Finally, the pure (> 99.5\%) electron and muon data samples~\footnote{The purity of these samples is provided by the SPS electron and muon beams.}  are used as validation sets.

In order to render the particle identification independent of the energy of the different species and thus to extend the method applied here to a larger scope than the beam purification, the pion samples of different energies are mixed before using the BDT technique. The same procedure is applied for muon and electron samples.

\section{Particle identification using Boosted Decision Trees}
Thanks to the high granularity of the SDHCAL, we can use the MVA methods to mine the information of the shape of electromagnetic and hadronic shower to classify muons, electrons and pions. The BDT method is one of the widely used MVA methods to perform such classification tasks. The BDT is a model that combines many less selective decision trees\footnote{ A decision tree takes a set of input variables and splits input data
recursively based on those variables.} into a strong classifier to achieve a much better performance.

\label{Para:BDT}
 \subsection{BDT input variables}

 The six variables we use to distinguish hadronic showers from electromagnetic showers and from muons are described below. A common right-handed coordinate system is used throughout the SDHCAL whose 48 layers were placed perpendicular to the incoming beams. The origin of the system is defined as the center of the first of the 48 SDHCAL's layers.    The $x$-$y$ plane is parallel to the SDHCAL layers and referred to as the transverse plane while the $z$-axis runs parallel to the incoming beam. 

\begin{figure}[tbp]
\begin{center}
\includegraphics[width=0.65\textwidth]{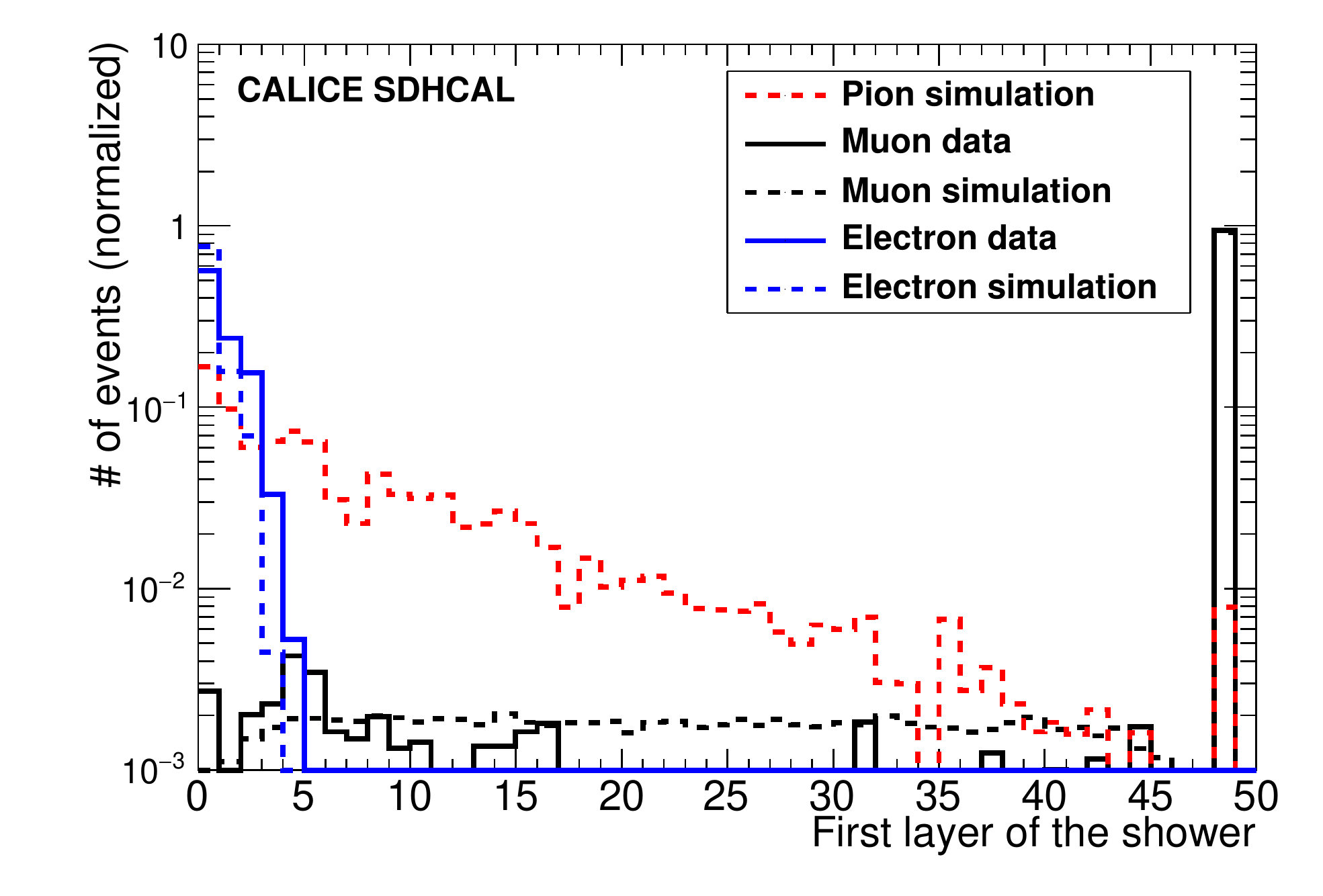}
\caption{\label{fig:a} Distribution of the first layer of the shower~(Begin). Layer 0 refers to the first layer of the prototype.  Continuous lines refer to data while dashed ones to the simulation. 
Layer 48 is the virtual layer after the last layer and used to tag events not fulfilling first layer criteria.}
\end{center}
\end{figure}

\begin{figure}[tbp]
\begin{center}
\includegraphics[width=0.65\textwidth]{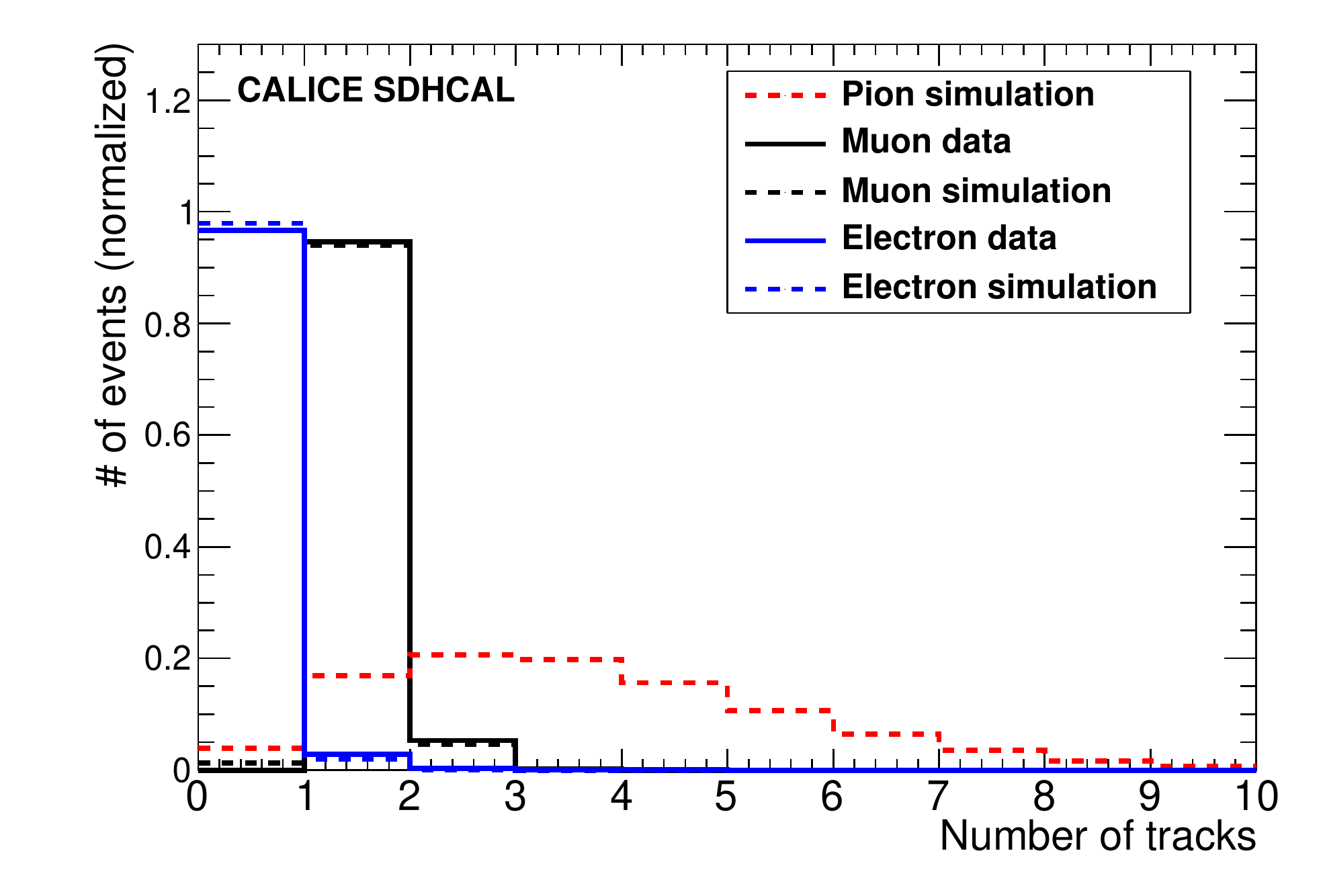}
\caption{\label{fig:b} Distribution of number of the tracks in the shower~(TrackMultiplicity).  Continuous lines refer to data while  dashed ones to the simulation.}
\end{center}
\end{figure}

\begin{figure}[tbp]
\begin{center}
\includegraphics[width=0.65\textwidth]{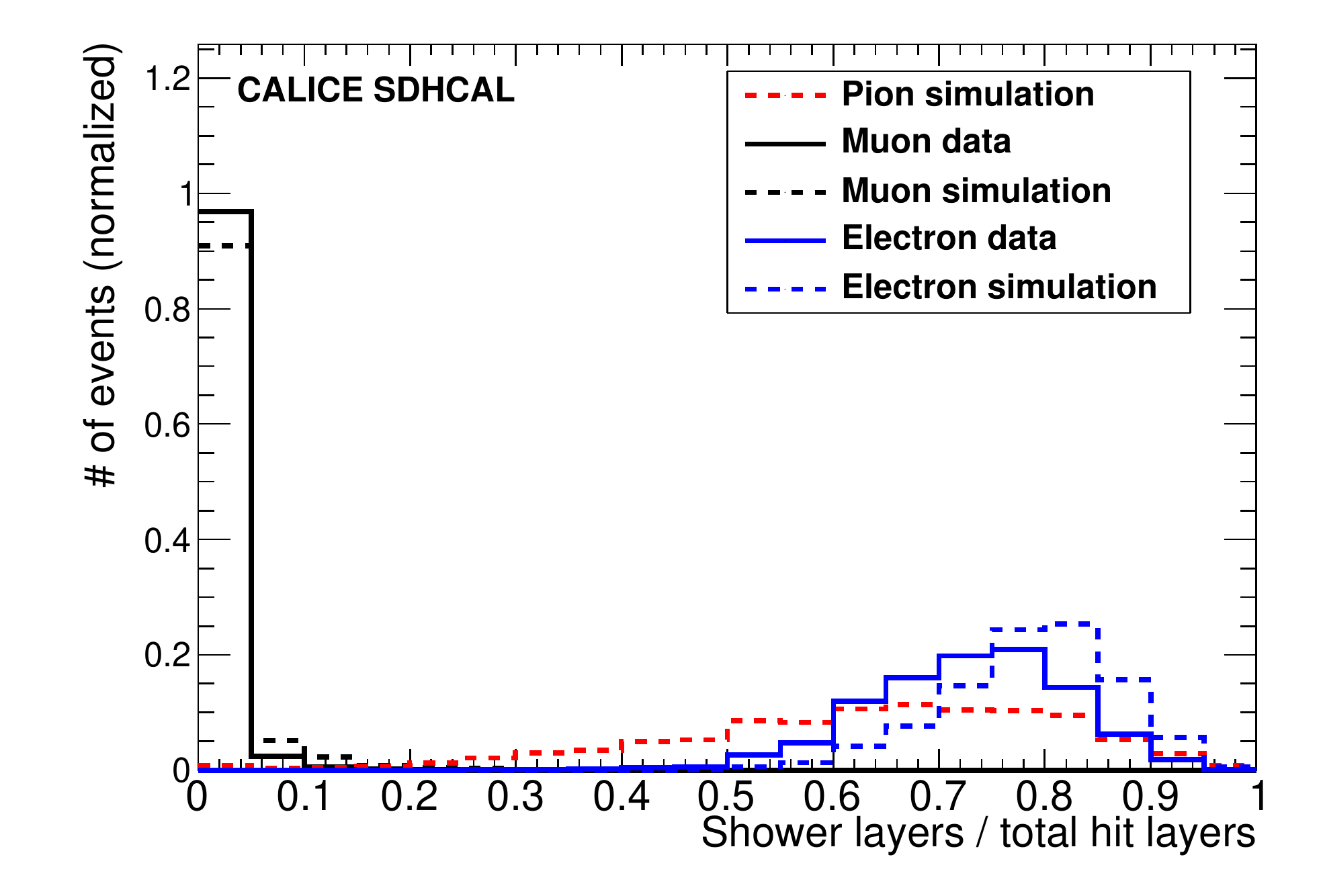}
\caption{\label{fig:c} Distribution of ratio of the number of layers  in which RMS of the hits' position in the $x$-$y$ plane exceeds 5~cm over the total number of fired layers~(NinteractingLayers/NLayers). Continuous lines refer to data while dashed ones to the simulation.}
\end{center}
\end{figure}

\begin{figure}[tbp]
\begin{center}
\includegraphics[width=0.65\textwidth]{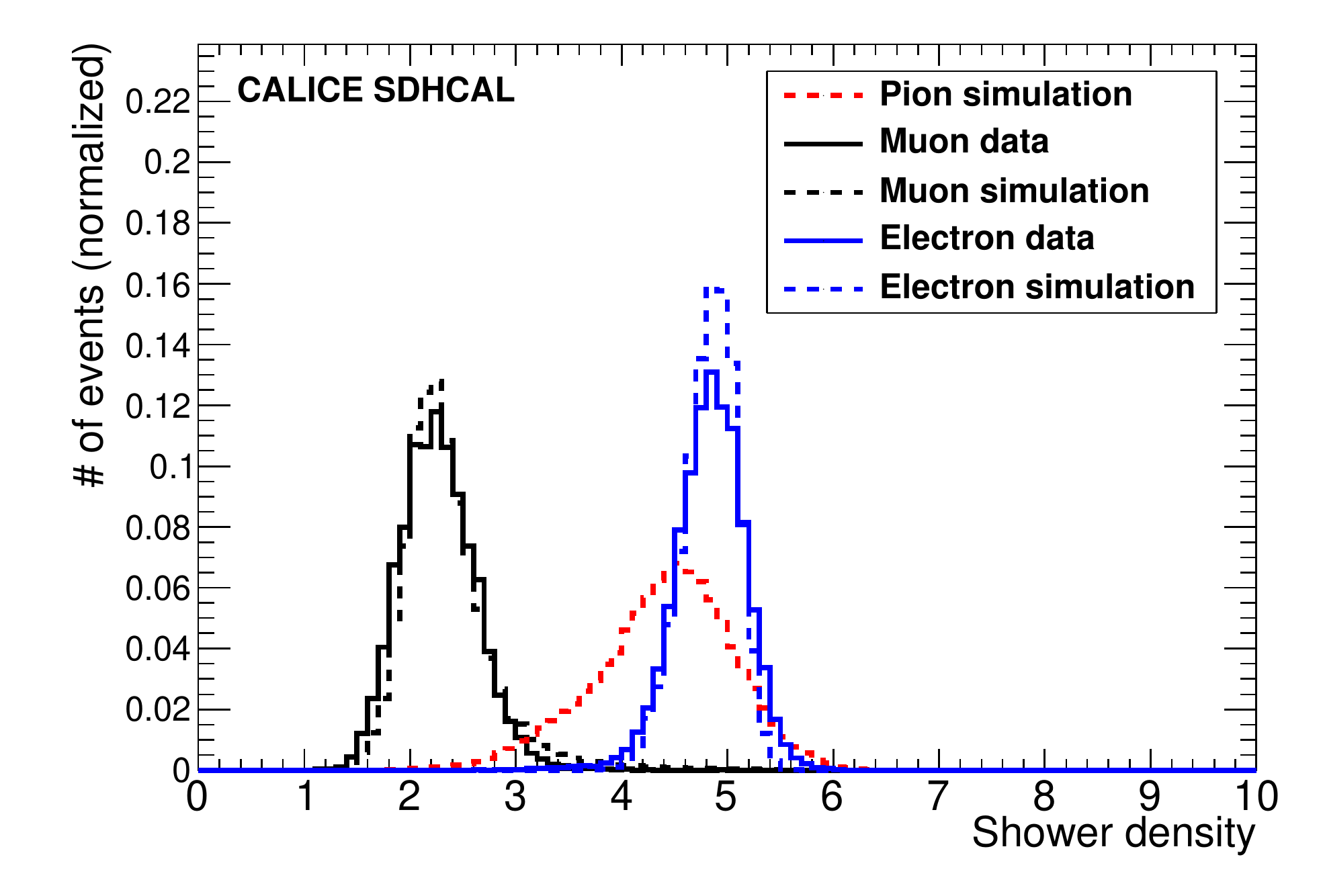}
\caption{\label{fig:d} Distribution of the average number of neighbouring hits surrounding one hit~(Density). Continuous lines refer to data while dashed ones to the simulation.}
\end{center}
\end{figure}

\begin{figure}[tbp]
\begin{center}
\includegraphics[width=0.65\textwidth]{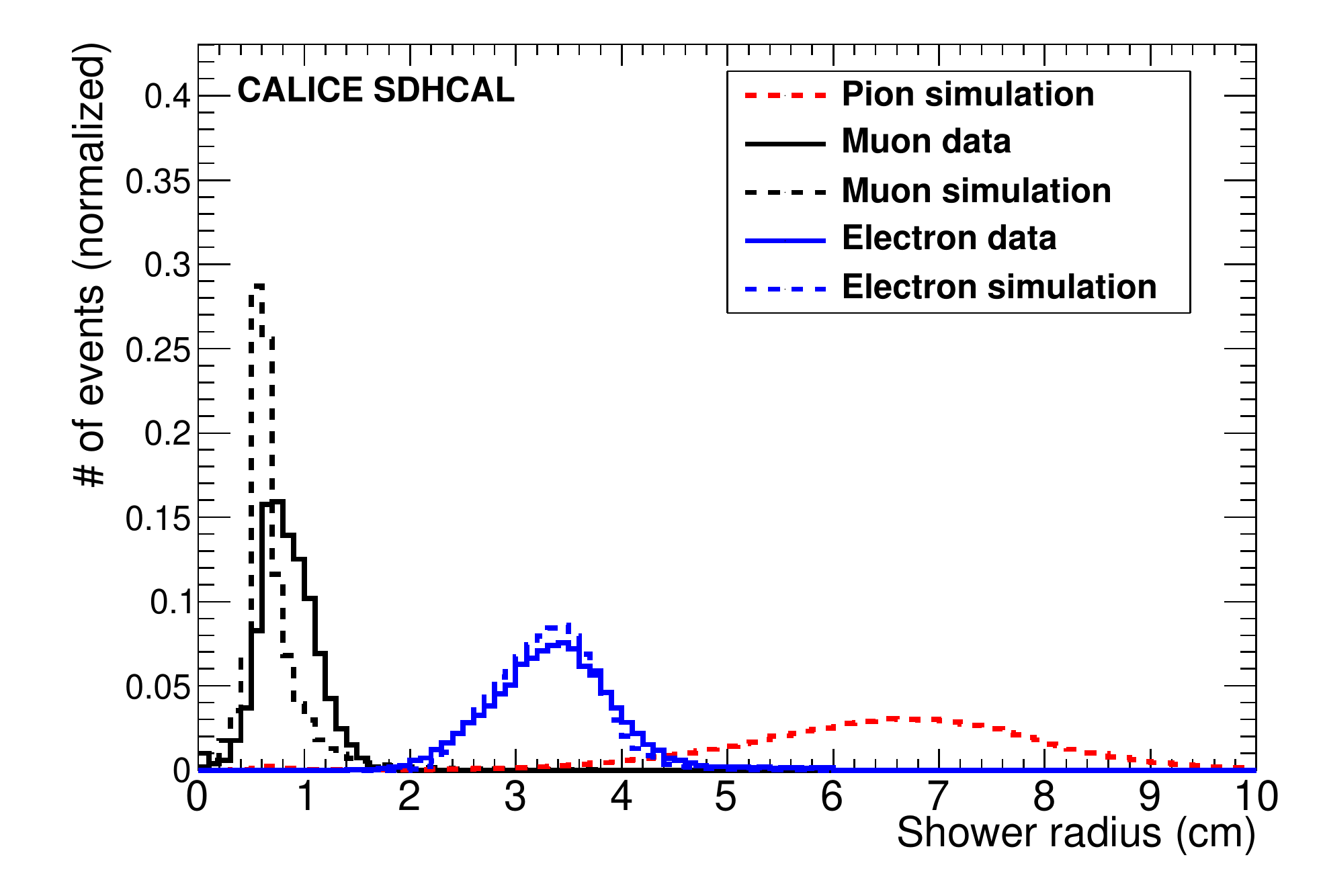}
\caption{\label{fig:e} Distribution of  the average radius of the shower~(Radius). Continuous lines refer to data while dashed ones to the simulation.}
\end{center}
\end{figure}

\begin{figure}[tbp]
\begin{center}
\includegraphics[width=0.65\textwidth]{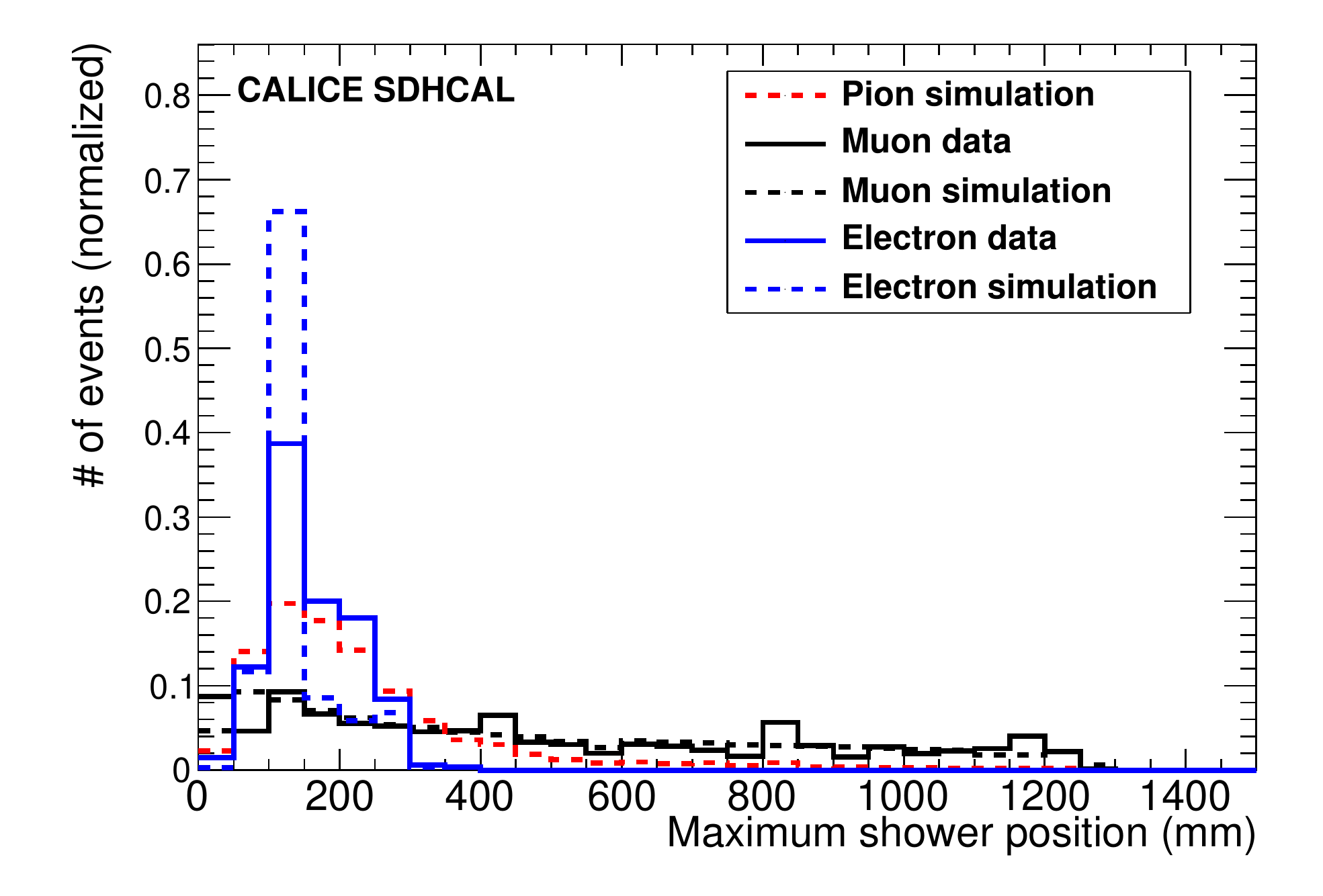}
\caption{\label{fig:f} Distribution of the position of the layer with the maximum radius~(Length). Continuous lines refer to data while dashed ones to the simulation.}
\end{center}
\end{figure}

 \begin{itemize}

\item {\bf  First layer of the shower (Begin) }:  The probability of a particle to interact in the calorimeter depends on  the particle nature and the calorimeter material properties. The distribution of the coordinate $z$ of the layer in which the first inelastic interaction takes place,  follows an exponential law. It is proportional  to $\exp{(-\frac{z}{X_0})}$ for electrons and to  $\exp{(-\frac{z}{\lambda_I})}$ for pions, where $X_0$ and $\lambda_I$ are effective radiation length and nuclear interaction length for the SDHCAL material composition, respectively.  To define the first layer in which the shower starts we look for the first layer along the incoming particle direction, which contains at least 4 fired pads. To eliminate fake shower starts due to accidental noise or a locally high multiplicity, the following 3 layers after the first one are also required to have more than 4 fired pads in each of them.   
         Particles crossing the calorimeter without interaction are assigned the value of  48, which is the case for most of the muons in the studied beam except the radiative ones.
        Figure~\ref{fig:a} shows the distribution of the first layer of the shower in the SDHCAL prototype for pions, electrons and muons as obtained from the simulation and data.

\item {\bf Number of tracks segments in the shower (TrackMultiplicity)}: Applying the Hough Transform~(HT) technique to single out the tracks in each event as described in Ref.~\cite{HT}, we estimate the number of tracks segments in the pion, electron and muon events. A HT-based segment candidate is considered as a track segment if there are more than 6 aligned hits with not more than one layer separating two consecutive hits.  Electron showers feature almost no track segment while most of the hadronic showers have at least one. For muons, except for some radiative muons, only one track is expected as can be seen in Fig.~\ref{fig:b}.

\item {\bf Ratio of shower layers over total hit layers (NinteractingLayers/NLayers)}: This is the ratio between the number of layers in which the Root Mean Square (RMS) of the hits' position in the $x$-$y$ plane exceeds 5~cm in both $x$ and $y$ directions and the total number of layers with at least one fired pad. It allows, as can be seen in Fig.~\ref{fig:c}, an easy discrimination of muons (even the radiative ones) from pions and electrons. It allows also a slight separation between pions and electrons.

\item {\bf Shower density (Density):}  This is the average number of the neighbouring hits located  in the $3\times 3$ pads around one of the hits including the hit itself in the given event. Figure~\ref{fig:d} shows clearly that electromagnetic showers are more compact than the hadronic showers as expected.

\item {\bf Shower radius (Radius):} This is the RMS of hits distance with respect to the event axis. To estimate the event axis, the average positions of the hits in each of the  ten first fired layers of an event are used to fit a straight line. The straight line is then used as the event axis. Figure~\ref{fig:e} shows the average radius of the three particle species in the SDHCAL.

\item {\bf Shower maximum position (Length):} This is the distance between the shower start and the layer where the maximum RMS of hit transverse coordimates with respect to shower axis is detected. The distribution of this variable for different particle species is shown in  Fig.~\ref{fig:f}.


\end{itemize}

 Before using the variables listed above as input to the BDT method, we check that the variables distributions in the simulation are in agreement with data for  the muon and  electron beams which are quite pure. Figures~\ref{fig:a}~-~\ref{fig:f} show that there is globally a good  agreement for the six variables of the two species even though the agreement is not perfect in particular for electrons. 

\subsection{The two approaches to build the BDT-based classifier}
  In order to take into account the small difference observed in some  variable distributions between data and simulation, and to cross-check the particle identification using the BDT method, we adopt two different training strategies for the BDT-based classifier. The first approach, referred to as MC Training, uses simulation samples of pions, electrons and muons as training sets. The second, referred to as Data Training, uses simulation samples of pions but electron and muon samples taken from data as training sets. 

  \subsubsection{MC Training Approach}
  The six variables of the simulated pion, muon and electron events  described in section 3.1 are used for the training and testing of the classifier. Events are chosen in alternating turns for the training and test samples as they occur in the source trees until the desired numbers of training and test events are selected. The training and test samples contain the same number of events for each event class.   Independent samples of signal events (pions) and of  the different background contributions (electron and muons) are used.
          The ratio between signal and  each background (electron or muon) events is 1 for training and test samples.  After the training, the BDT provides the relative weight of each variable as a measure of  distinguishing  signal from background. Two BDT-based classifiers are proposed here. The first~(BDT$_{\pi\mu}$)~is used to discriminate pions against muons and the second~(BDT$_{\pi e}$)~to discriminate against electrons. 
Table~\ref{tab:i} shows the variable ranking according to their separation power in the BDT$_{\pi\mu}$ while Tab.~\ref{tab:ii} gives their separation power in the case of BDT$_{\pi e}$.  The BDT algorithm using the variables and their respective weights is then applied to the test samples.  The output of the BDT applied to  each of the test sample events is a variable belonging to the interval  [-1,1]  with the positive value representing more signal-like events and the negative more background-like events.

\begin{table}[tbp]
\centering
\caption{\label{tab:i} Variable ranking of separation power in the case of BDT$_{\pi\mu}$. }
\begin{tabular}{l||l}
\hline
\multirow{2}{*}{Rank : Variable} &\multirow{2}{*}{Variable relative weight}\\
 & \\
 \hline
1 : Length &0.233\\
2 : Density & 0.225\\
3 : NInteractinglayer$/$Nlayer  & 0.163\\
4 : Radius & 0.160 \\
5 : Begin & 0.139\\
6 : TrackMultiplicity & 0.080\\
\hline
\end{tabular}
\end{table}

\begin{table}[tbp]
\centering
\caption{\label{tab:ii} Variable ranking of separation power in the case of BDT$_{\pi e}$.}
\begin{tabular}{l||l}
\hline
\multirow{2}{*}{Rank : Variable} &\multirow{2}{*}{Variable relative weight}\\
 & \\
 \hline
1 : Radius & 0.204\\
2 : NInteractinglayer$/$Nlayer  & 0.203\\
3 : Density &0.194\\
4 : Length & 0.151\\
5 : Begin & 0.145\\
6 : TrackMultiplicity & 0.101\\
\hline
\end{tabular}
\end{table}

\begin{figure}[tbp]
\begin{center}
\begin{tabular}{cc}
\includegraphics[width=0.49\textwidth]{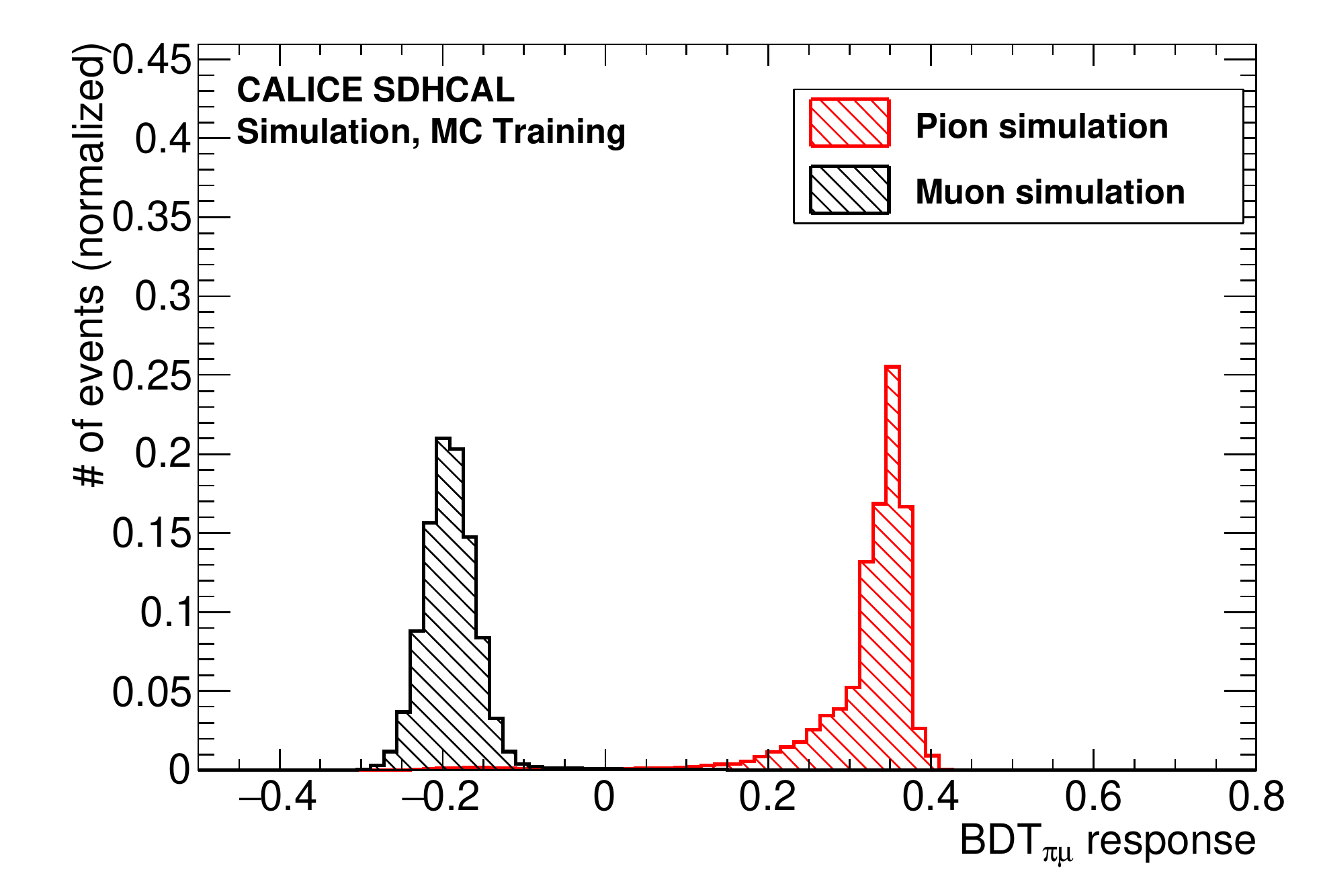} &
\includegraphics[width=0.49\textwidth]{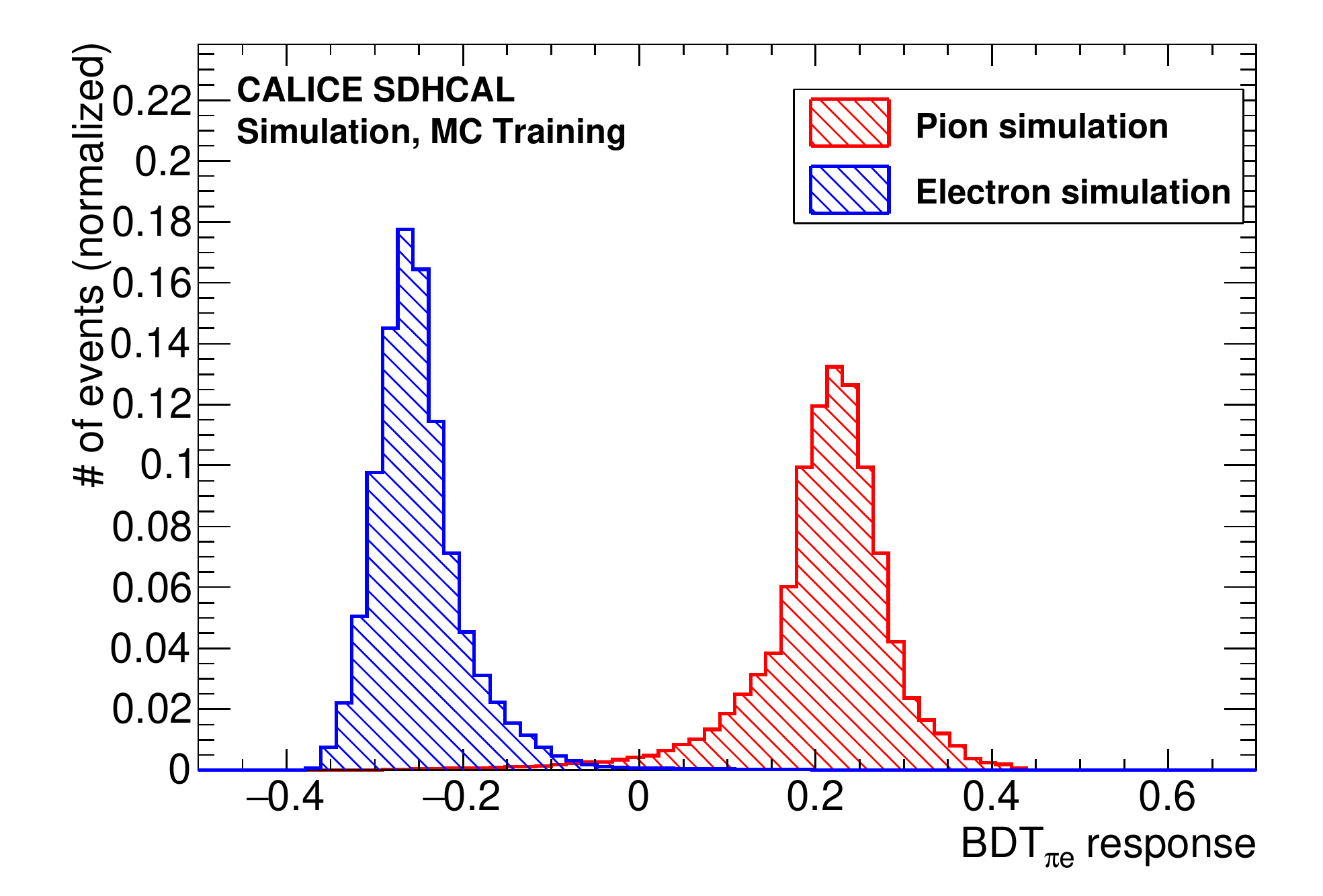} \\
\end{tabular}
\caption{The BDT output of the BDT$_{\pi\mu}$~(left)~and BDT$_{\pi e}$~(right)~built with simulation samples.}
\label{fig:BDT_MCtraining}
\end{center}
\end{figure}

\begin{figure}[tbp]
\begin{center}
\begin{tabular}{cc}
\includegraphics[width=0.49\textwidth]{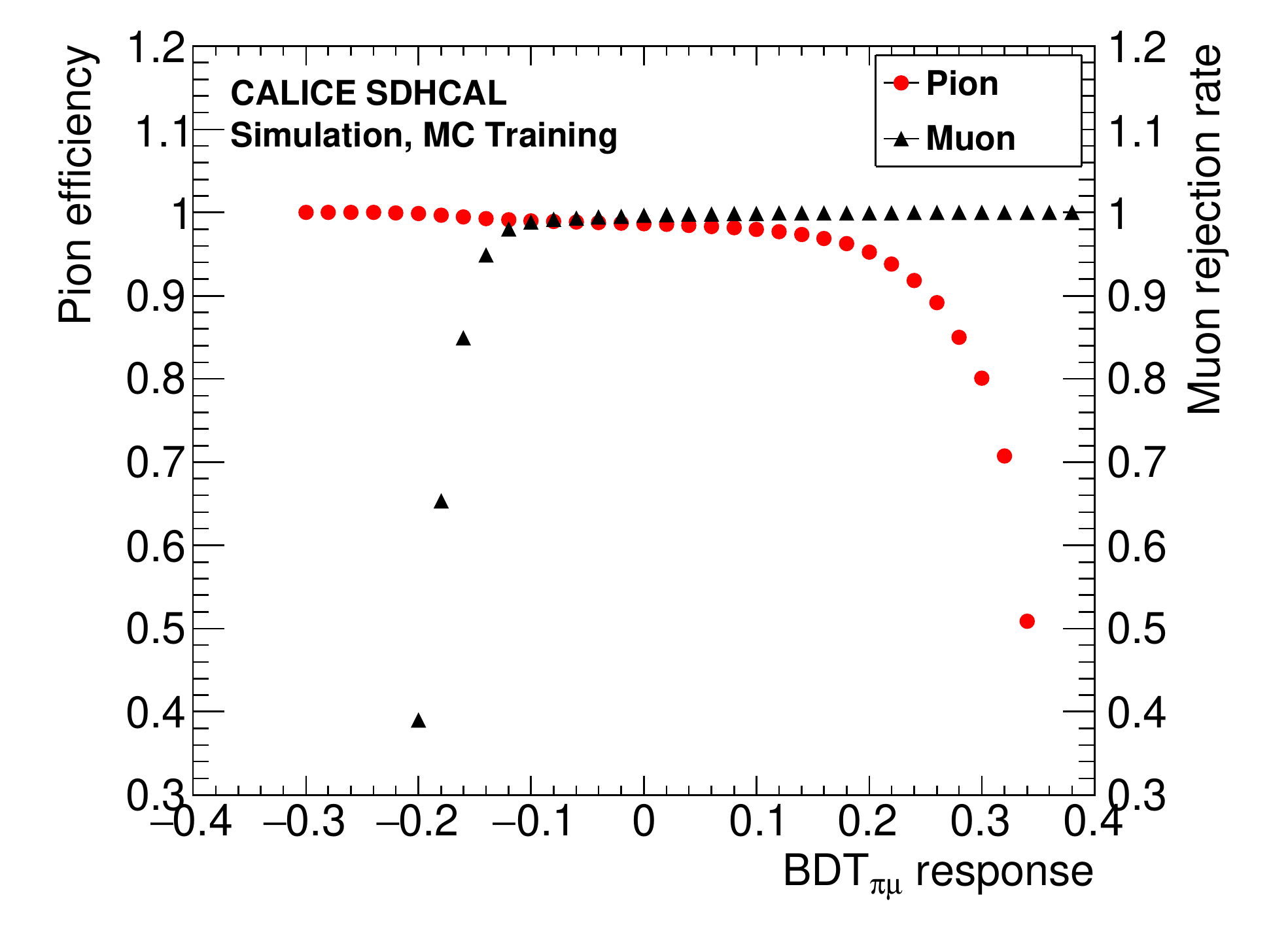} &
\includegraphics[width=0.49\textwidth]{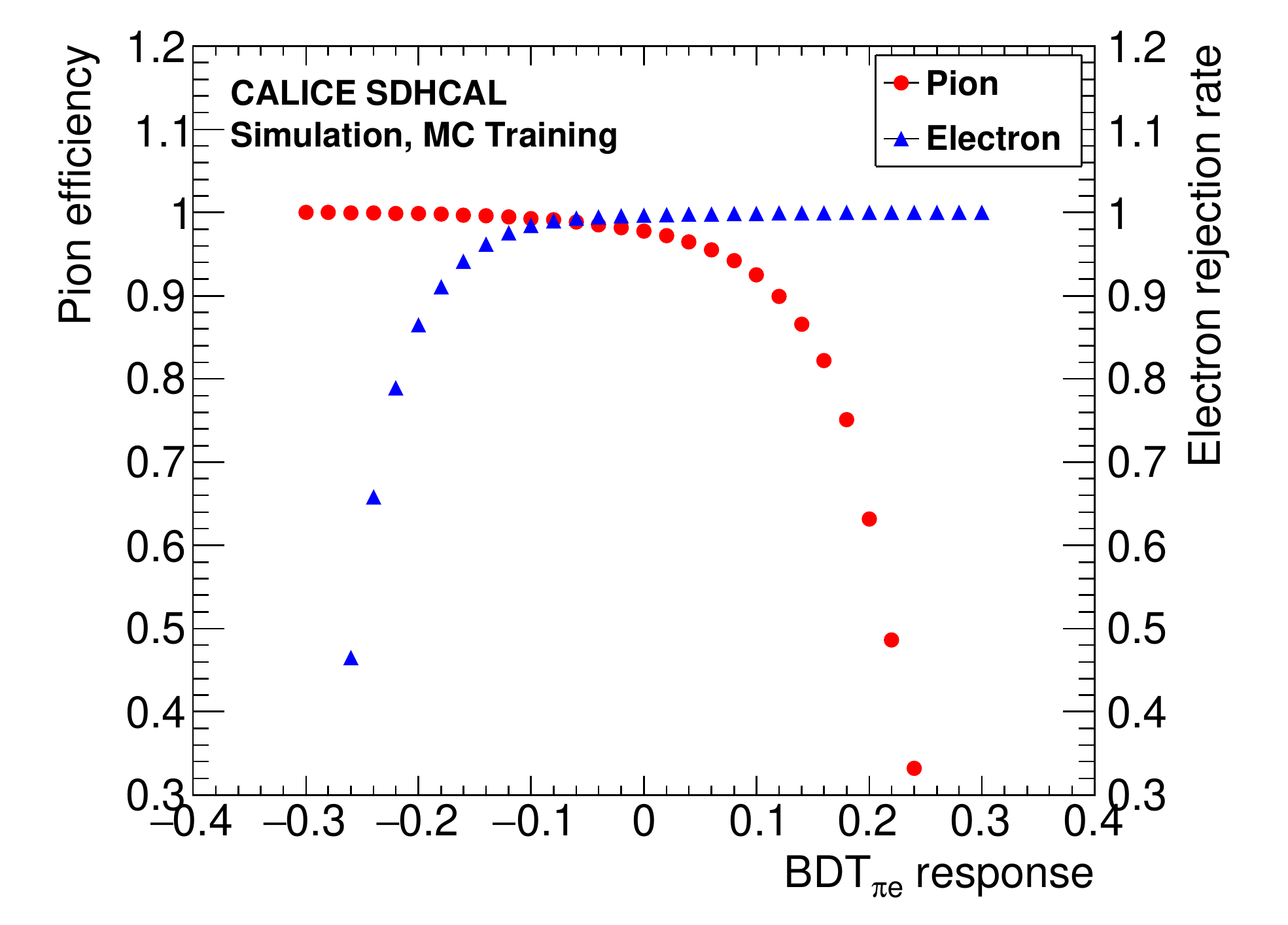} \\
\end{tabular}
\caption{Pion efficiency and muon rejection rate~(left) and pion efficiency and electron rejection rate~(right)  as a function of the BDT output.}
\label{fig:eff_MCtraining}
\end{center}
\end{figure}

\begin{figure}[tbp]
\begin{center}
\begin{tabular}{cc}
\includegraphics[width=0.49\textwidth]{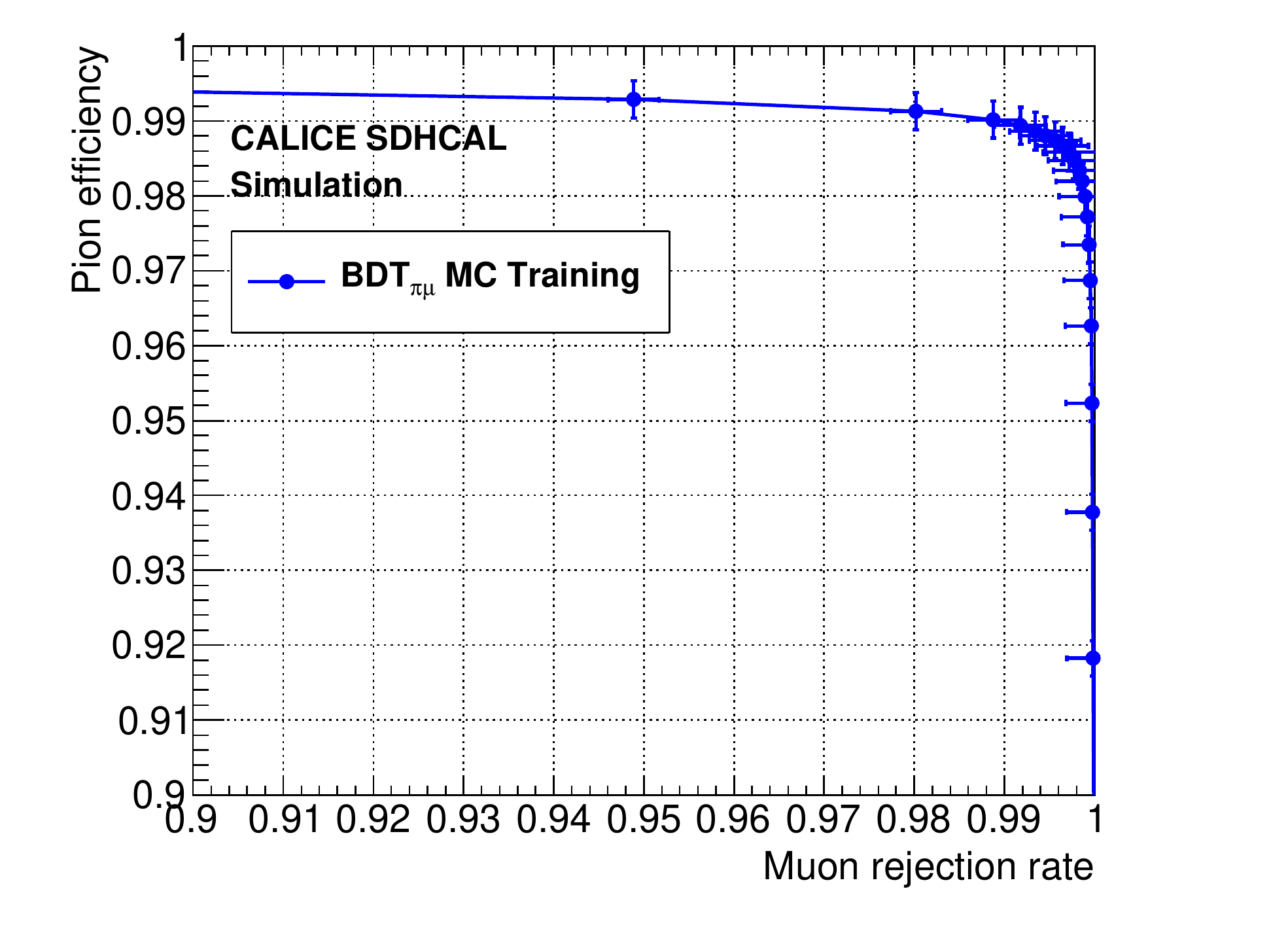} &
\includegraphics[width=0.49\textwidth]{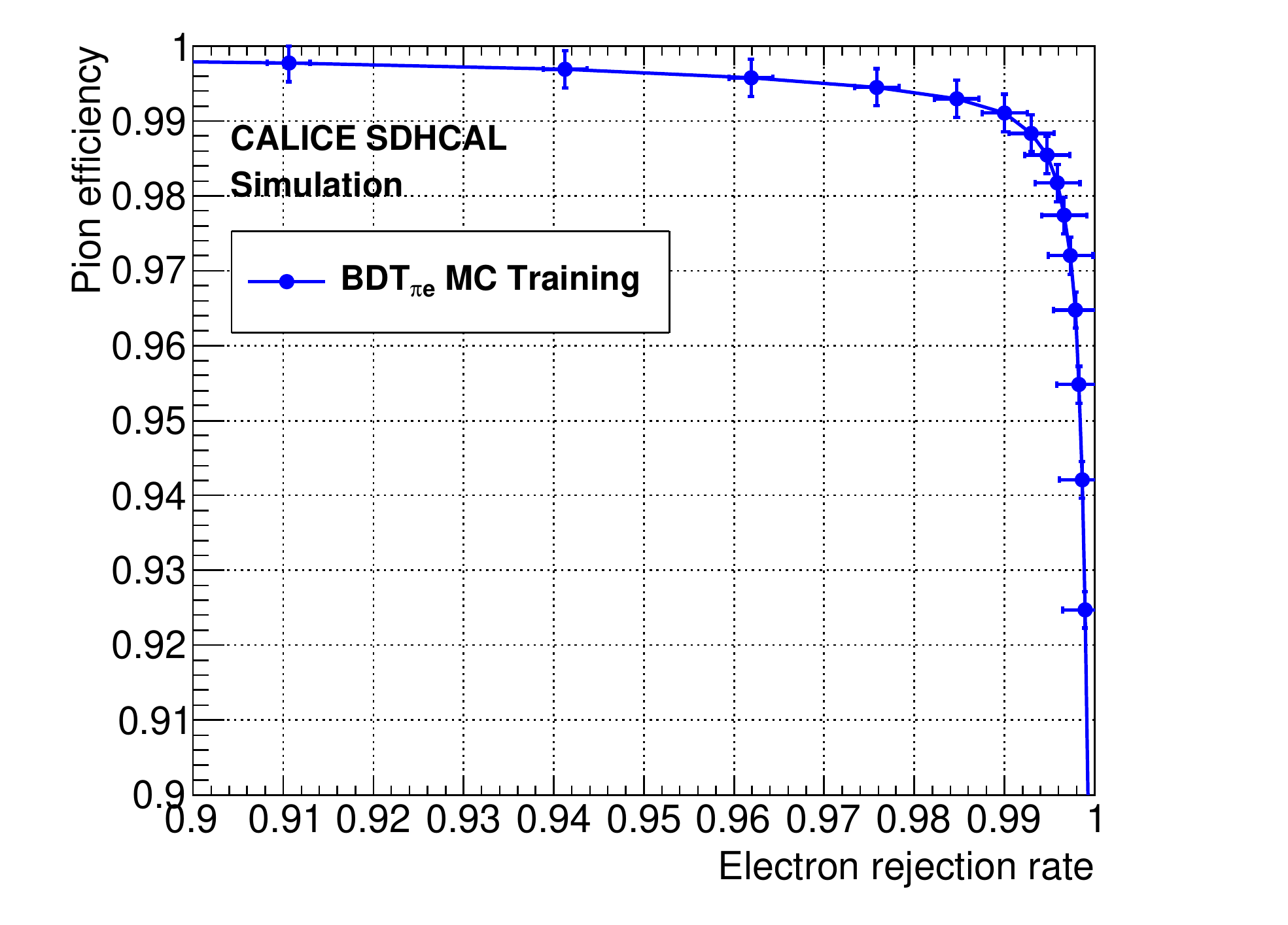} \\
\end{tabular}
\caption{Pion efficiency versus muon rejection rate(left) and pion efficiency versus electron rejection rate~(right).}
\label{fig:eff_MCtraining1}
\end{center}
\end{figure}

         Figure~\ref{fig:BDT_MCtraining}~(left)  shows the output of the BDT for a test sample made of pions and muons while  Fig.~\ref{fig:BDT_MCtraining}~(right) shows the output for a test sample made of pions and electrons.  The values differ significantly for signal and background suggesting thus a large separation power of the BDT approach. This is confirmed by Fig.~\ref{fig:eff_MCtraining}. The pion selection efficiency versus the muon~(electron) rejection of the test sample is shown in Fig.~\ref{fig:eff_MCtraining1}~(left) and Fig.~\ref{fig:eff_MCtraining1}~(right), respectively. A pion selection efficiency exceeding 99\% with a  muon and electron rejection of the same level~(> 99\%)  can be achieved. 

\begin{figure}[tbp]
\begin{center}
\begin{tabular}{cc}
\includegraphics[width=0.49\textwidth]{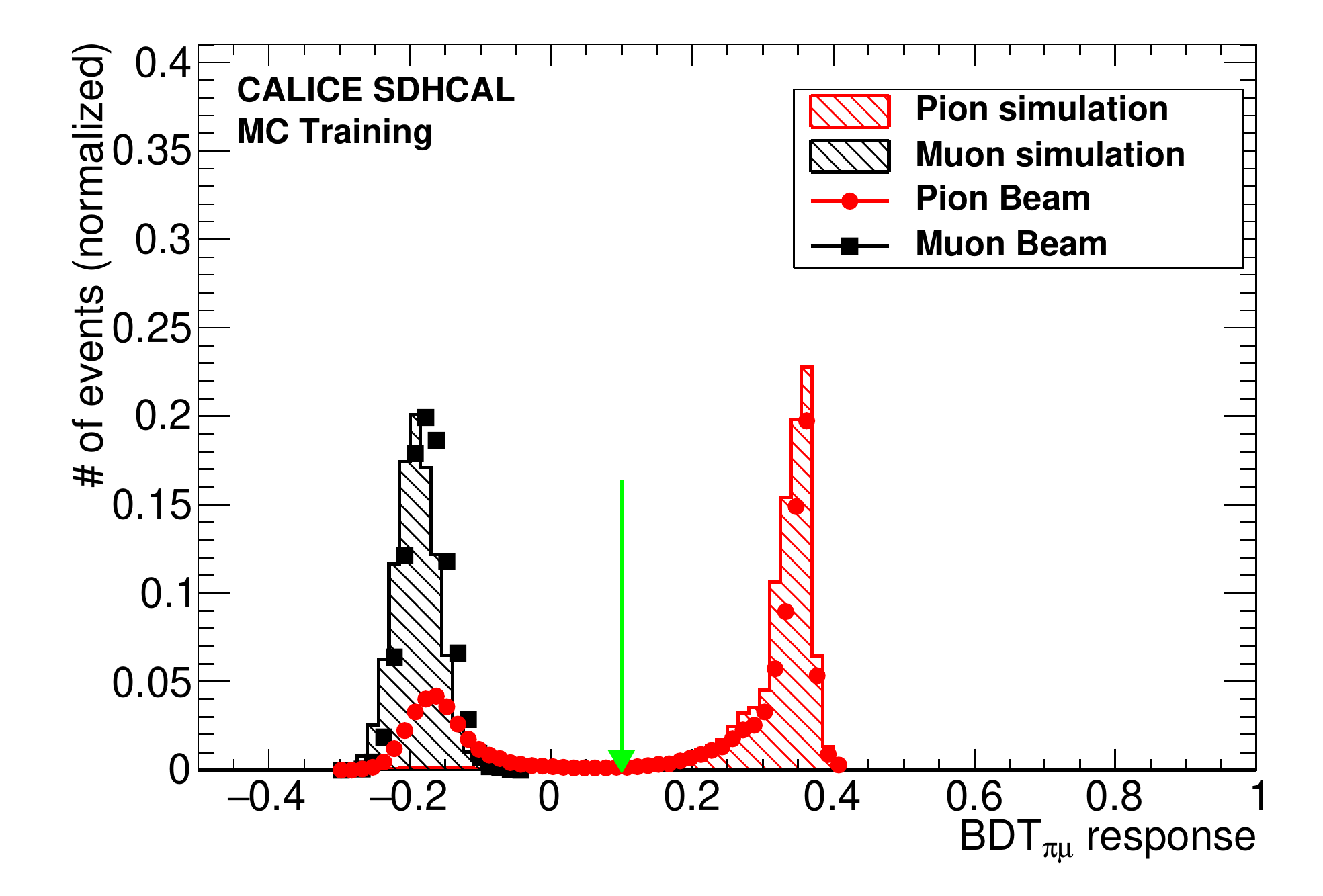} &
\includegraphics[width=0.49\textwidth]{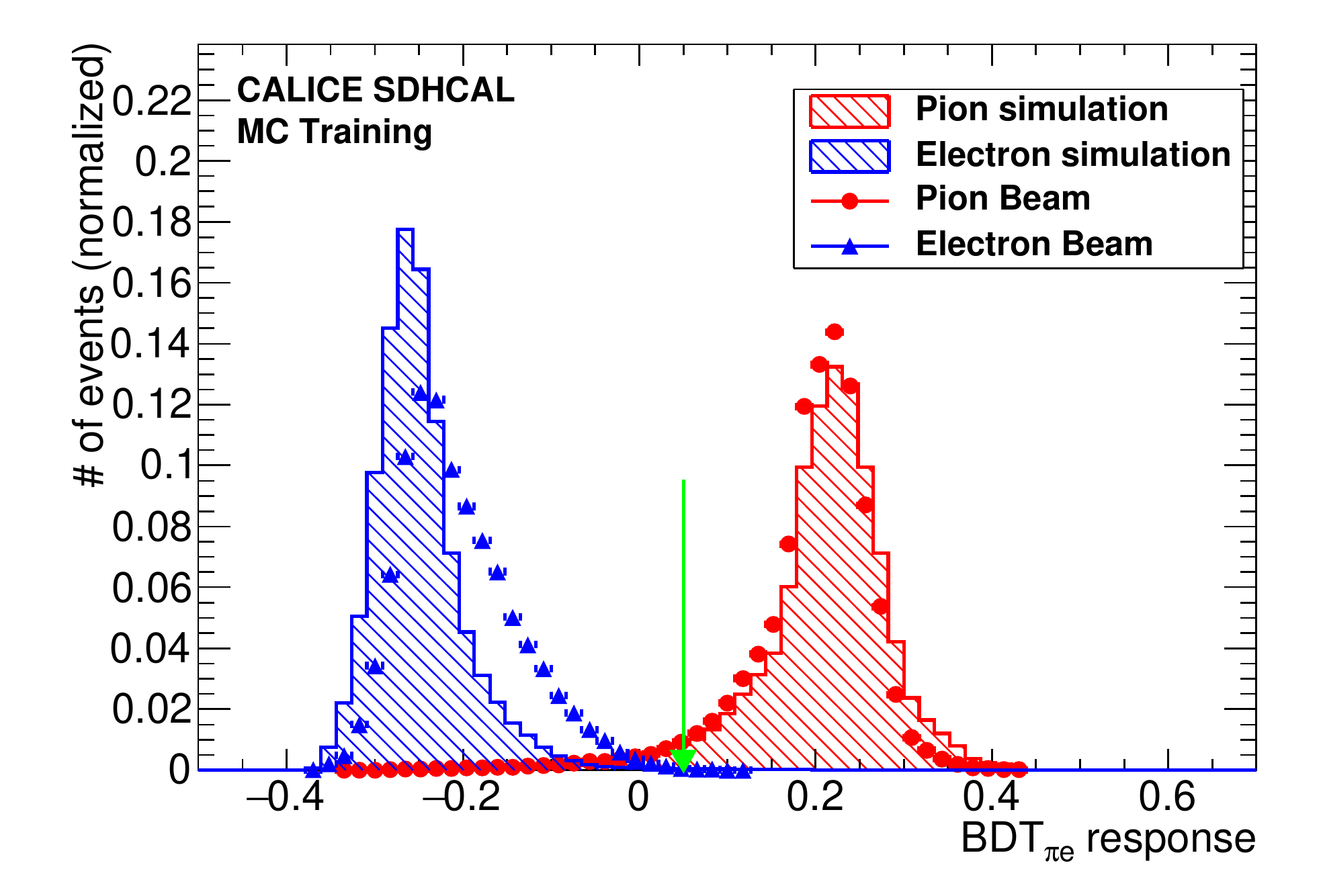} \\
\end{tabular}
\caption{The BDT output after using the BDT$_{\pi\mu}$~on the data pion sample (left) and the BDT output after using the BDT$_{\pi e}$~ on the same data pion sample after classified by BDT$_{\pi\mu}$ ~(right). A green arrow is shown on both  to indicate the BDT cut applied to clean the pion samples.}
\label{fig:BDT_MCtraining1}
\end{center}
\end{figure}

  In order to check the validity of these two classifiers, we use the pure beam samples of muons and electrons. Figure~\ref{fig:BDT_MCtraining1}~(left) shows the BDT output of BDT$_{\pi\mu}$ and Fig.~\ref{fig:BDT_MCtraining1}~(right) shows the case of BDT$_{\pi e}$. Beam muon results show a good agreement with respect to the simulated events.  A slight shift of the beam electron shape is observed with respect to the one obtained from the simulated events. This difference is most probably due to the fact that the distribution of  some variables in data and in the simulation are not identical. Next, as a first step of purifying the collected hadronic data events we apply the pion-muon classifier.  Figure~\ref{fig:BDT_MCtraining1} (left) shows the BDT$_{\pi\mu}$ response applied to the collected hadron events in the SDHCAL. We can clearly see that there are two maxima. One maximum in the muon range corresponds to the muon contamination of pion data and another one in the pion range.  Hence, to ensure the rejection of the muons in the sample, the BDT variable is required to be > 0.1.  The second step is to apply the  BDT$_{\pi e}$ to the remaining of the pion sample.  Figure~\ref{fig:BDT_MCtraining1} (right) shows the BDT$_{\pi e}$ output.  In order to eliminate the maximum of the electrons contamination and get almost a pure~( > 99.5\%) pion sample with limited loss of pion events, we apply to the pion samples a BDT$_{\pi e}$ cut of 0.05.

 \subsubsection{Data Training Approach}
\begin{figure}[tbp]
\begin{center}
\begin{tabular}{cc}
\includegraphics[width=0.49\textwidth]{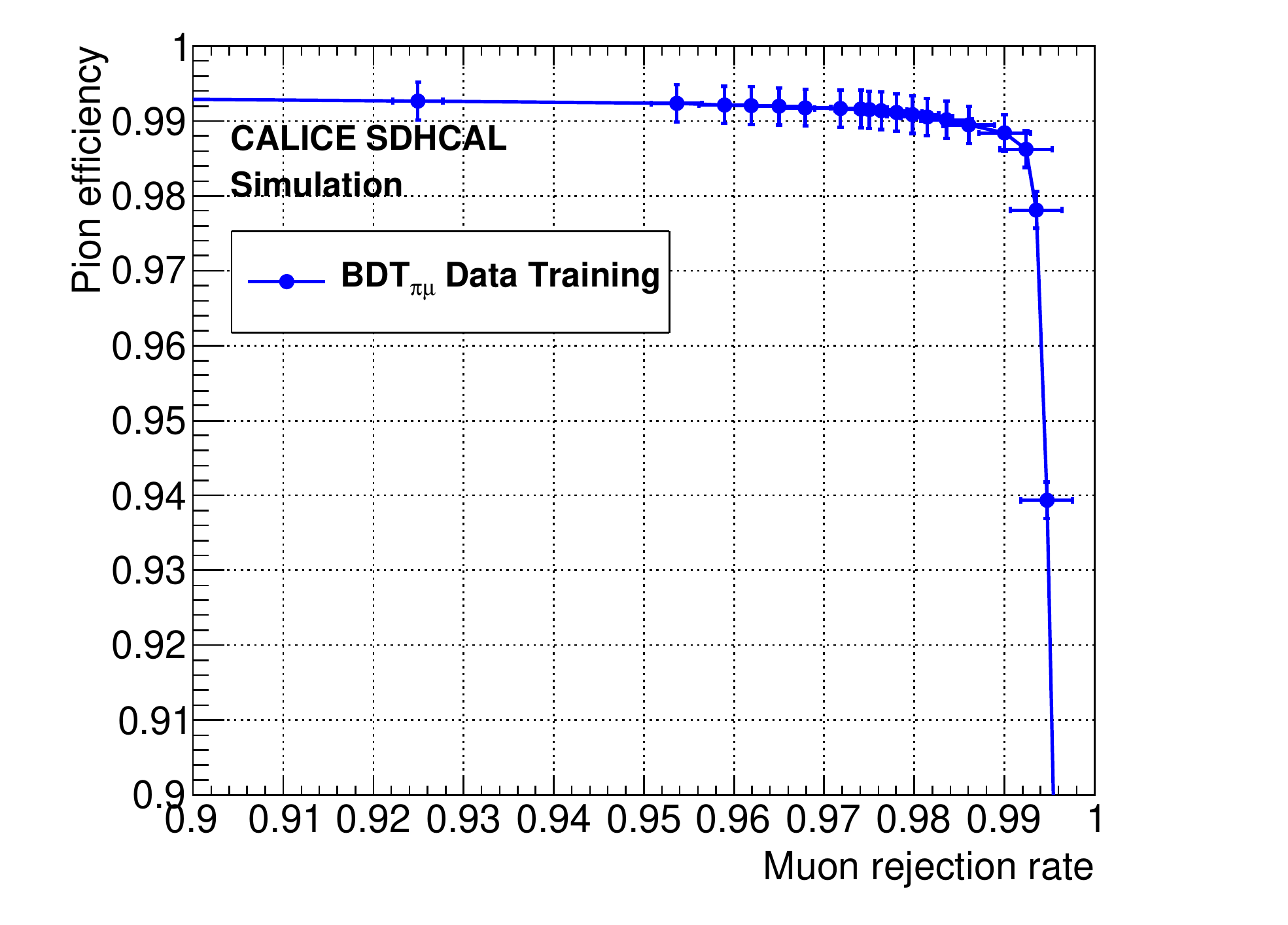}&
\includegraphics[width=0.49\textwidth]{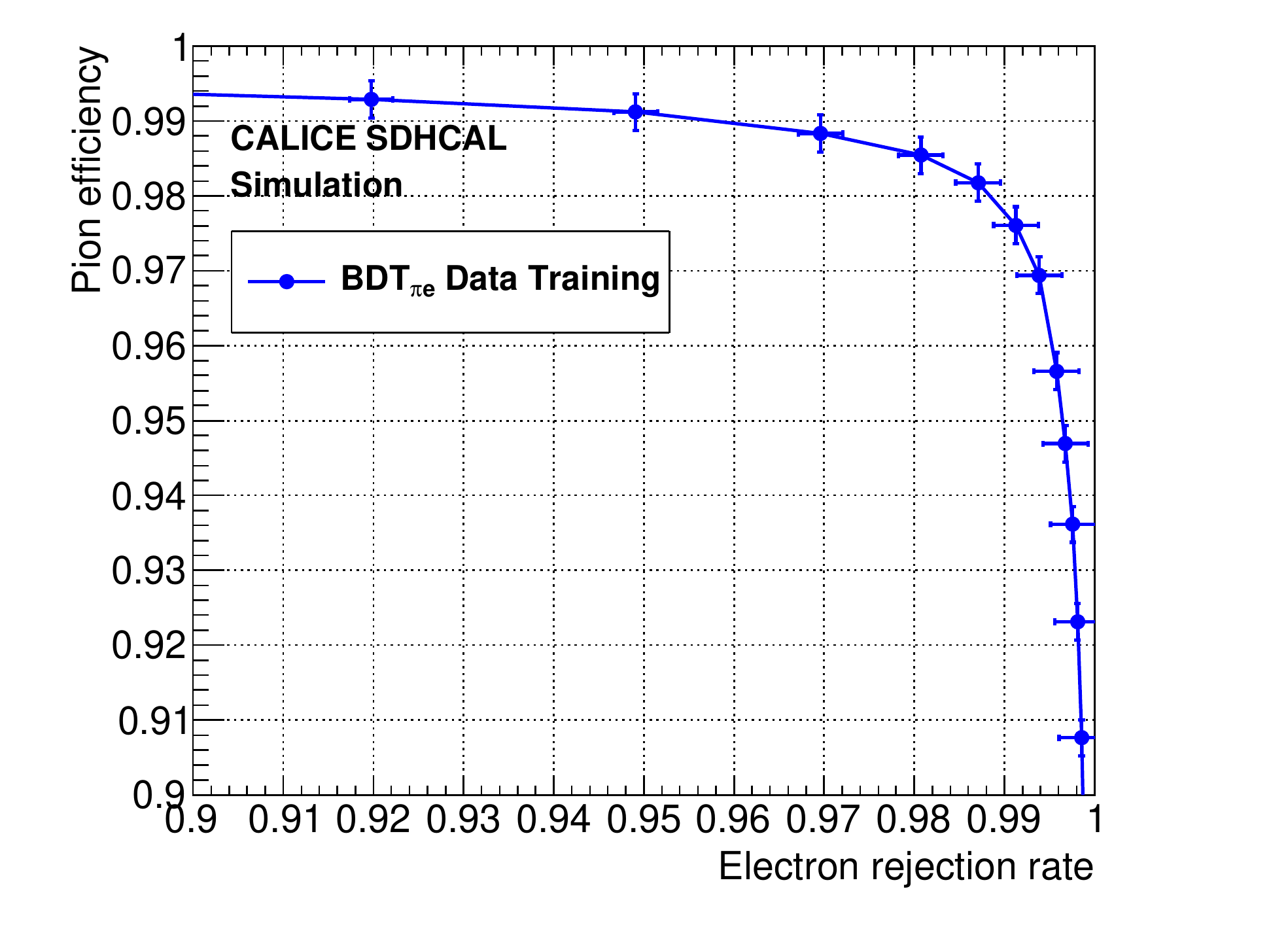}\\
\end{tabular}
\caption{Pion efficiency versus muon rejection rate~(left) and pion efficiency versus electron rejection rate~(right).}
\label{fig:eff_datatraining}
\end{center}
\end{figure}

\begin{figure}[tbp]
\begin{center}
\begin{tabular}{cc}
\includegraphics[width=0.49\textwidth]{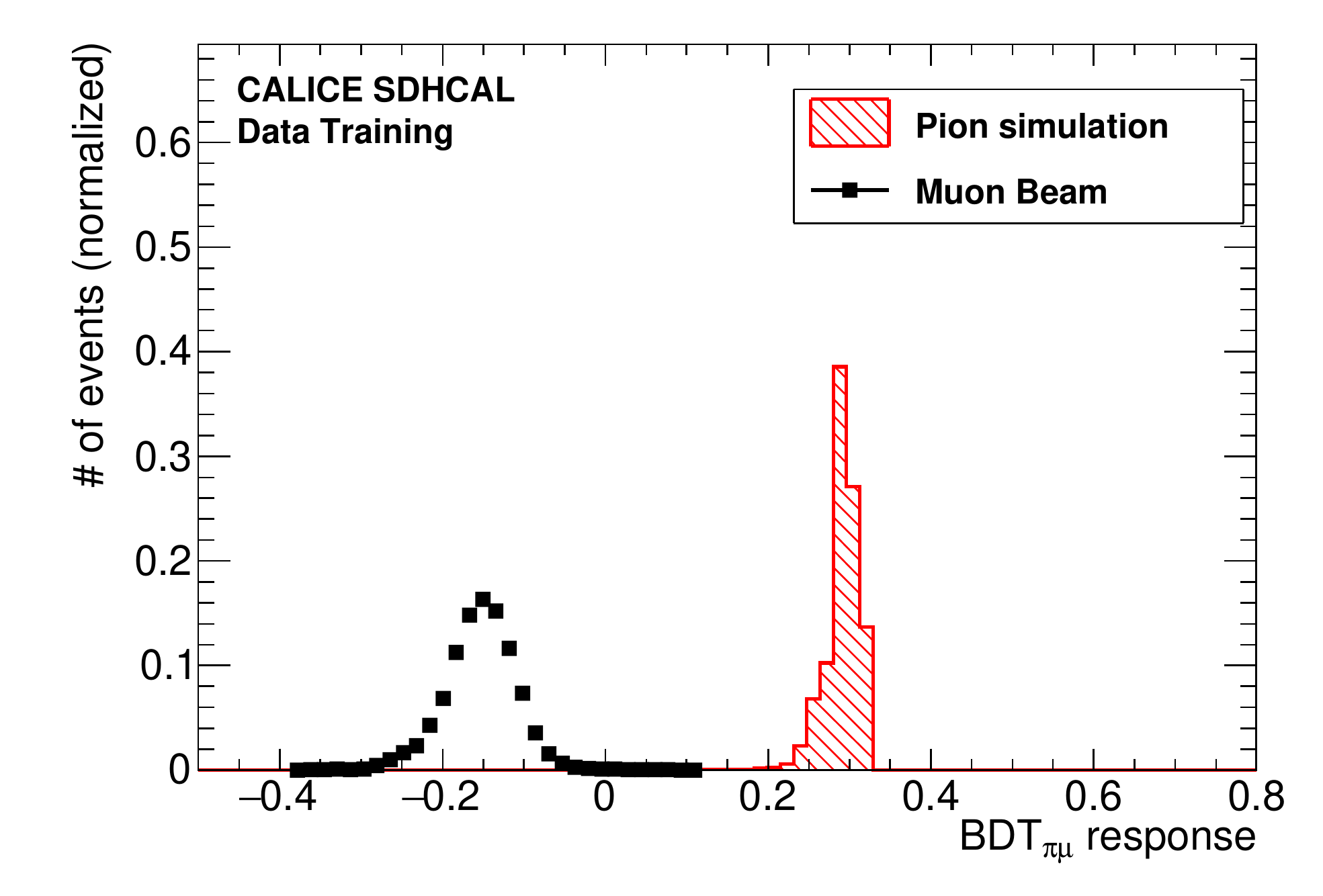}&
\includegraphics[width=0.49\textwidth]{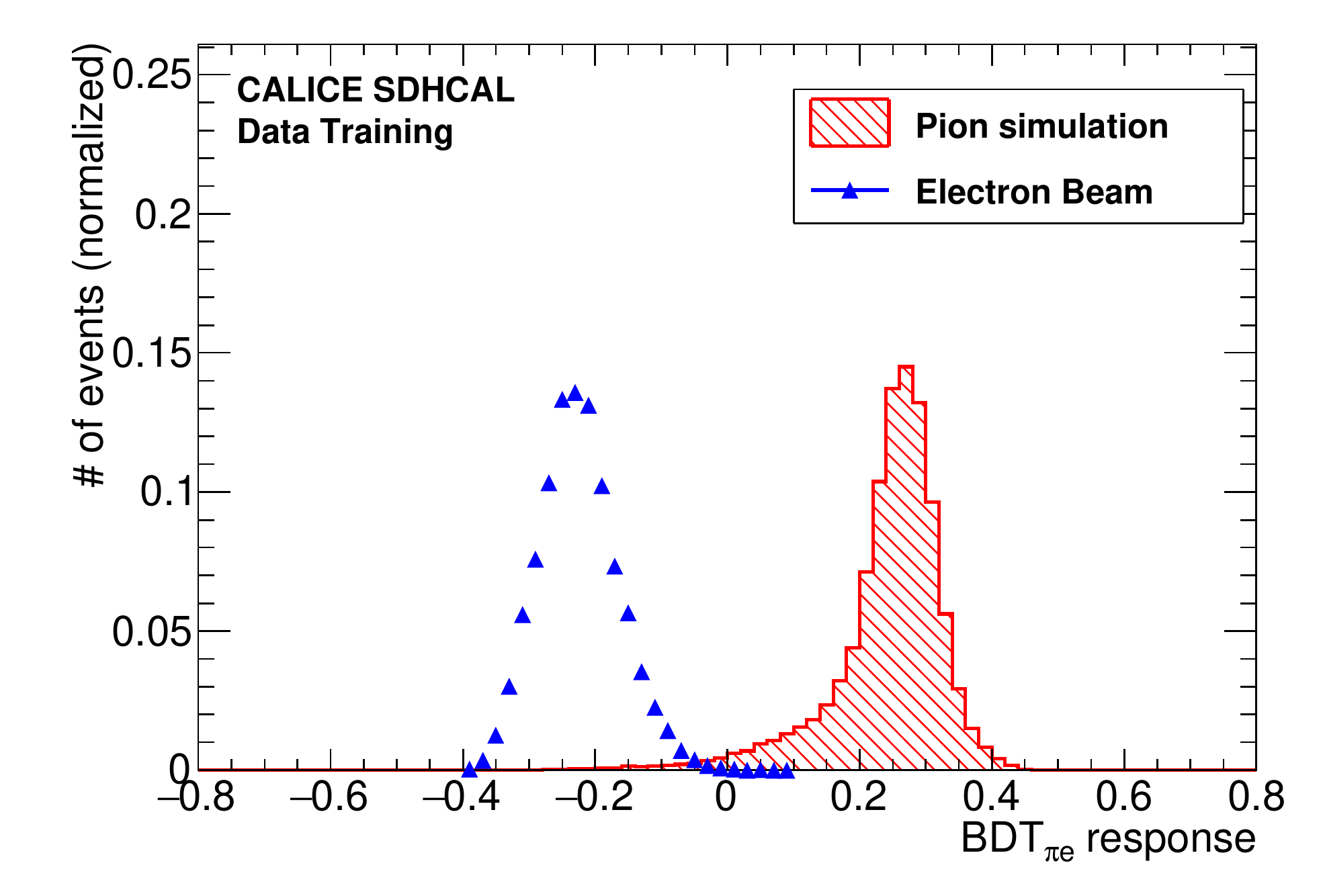}\\
\end{tabular}
\caption{BDT output of the BDT$_{\pi\mu}$~ built with pure beam  muons and  simulated pion samples (left) and of the  BDT$_{\pi e}$~ built with  pure beam electrons and simulated pion samples~(right)}
\label{fig:BDT_datatraining}
\end{center}
\end{figure}

\begin{figure}[tbp]
\begin{center}
\begin{tabular}{cc}
\includegraphics[width=0.49\textwidth]{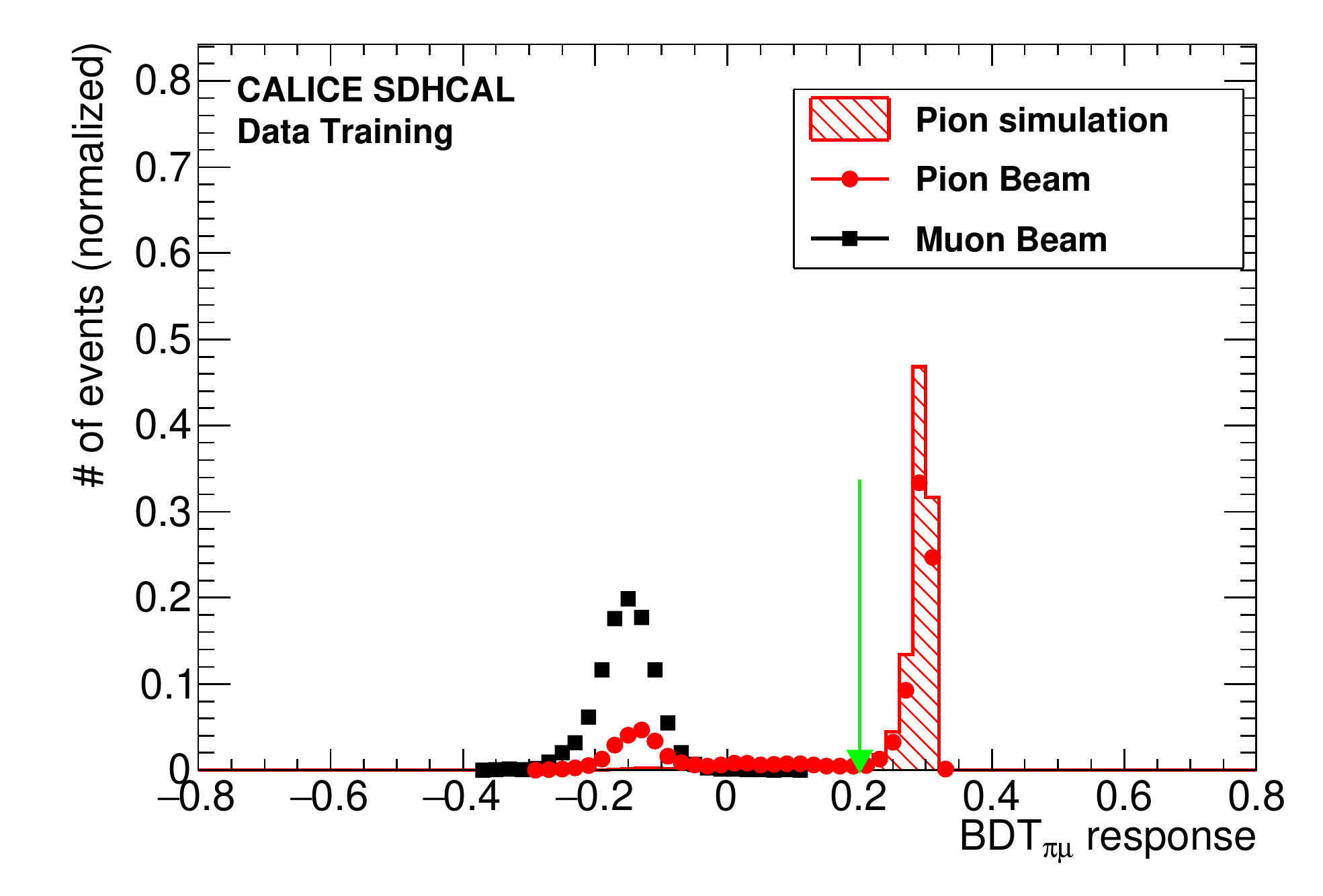}&
\includegraphics[width=0.49\textwidth]{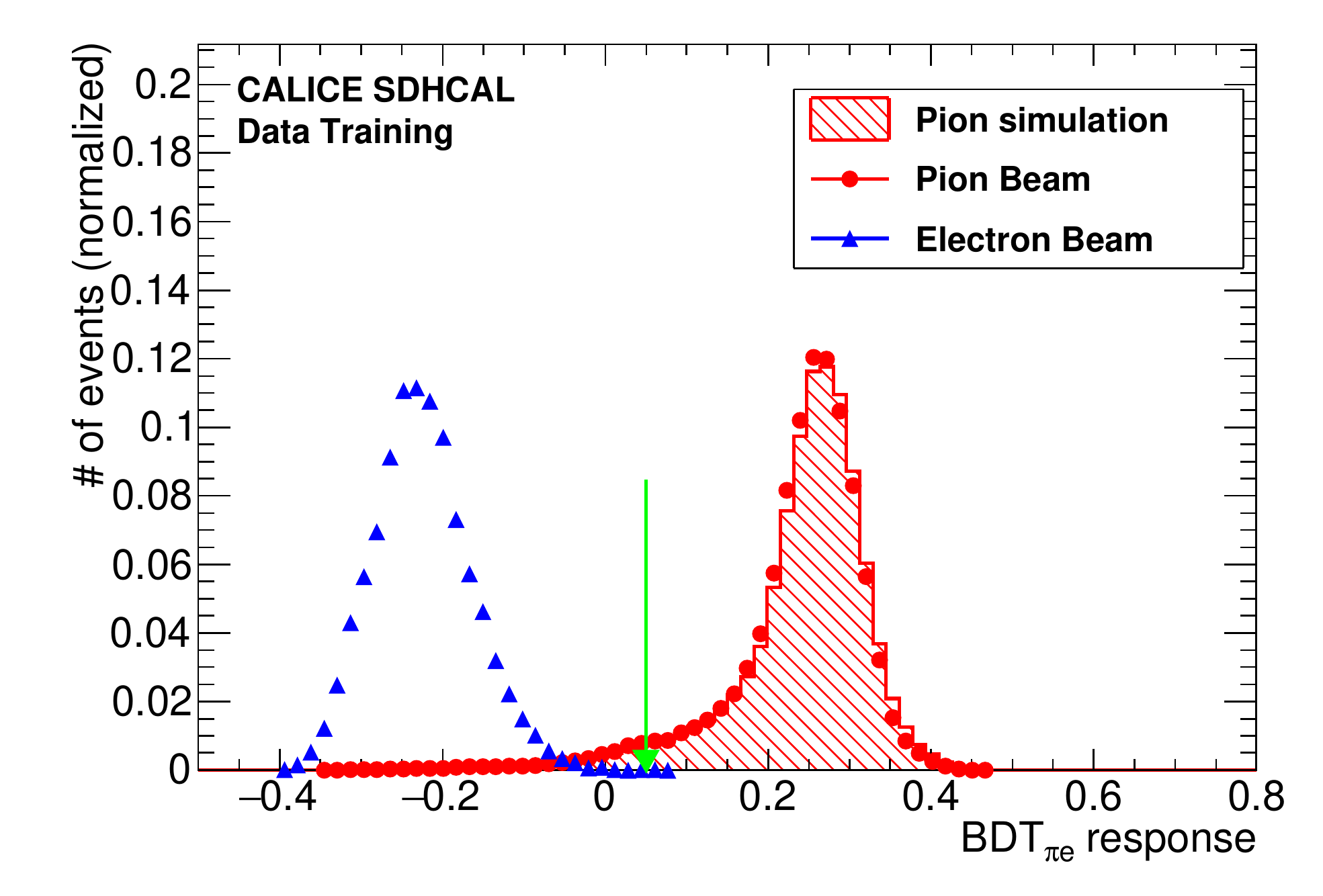}\\
\end{tabular}
\caption{The BDT output after using the BDT$_{\pi\mu}$~ on the data pion sample (left) and the BDT output after using the BDT$_{\pi e}$~ on the same pion sample after classified by BDT$_{\pi\mu}$ ~(right). A green arrow is shown on both  to indicate the BDT cut applied to clean the pion samples.}
\label{fig:BDT_datatraining_beam}
\end{center}
\end{figure}

 We use the same variables of the  MC Training approach on the data samples of muons and electrons  but  still on the  simulated pion samples to build two classifiers. Then we apply the same procedure as the MC Training approach.  Table~\ref{tab:rank_pimu_datatraining} and ~\ref{tab:rank_pie_datatraining} show the corresponding variables ranking for BDT$_{\pi\mu}$ and BDT$_{\pi e}$ according to their power separation importance. The difference  of variables weights of these two tables with respect to those obtained with MC training approach is explained by the slight difference of some variables distributions  between data and simulation.  Figure~\ref{fig:eff_datatraining} left~(right) gives the results of pion efficiency and muon~(electron) rejection rate. This shows that these two classifiers have very good pion efficiency and high background rejection rate. The left~(right) plot of Fig.~\ref{fig:BDT_datatraining} shows the BDT output of the BDT$_{\pi\mu}$~(BDT$_{\pi e}$). Clearly these two classifiers have very good separation power. We apply these classifiers to the raw pion beam samples. The results can be seen in Fig.~\ref{fig:BDT_datatraining_beam}. We apply a BDT cut value of 0.2 in the pion-muon separation stage and then a BDT cut value of 0.05 in the pion-electron separation stage.

\begin{table}[tbp]
\centering
\caption{\label{tab:rank_pimu_datatraining} Variable ranking of separation importance in the case of~BDT$_{\pi\mu}$.}
\begin{tabular}{l||l}
\hline
\multirow{2}{*}{Rank : Variable} &\multirow{2}{*}{Variable relative weight}\\
 & \\
 \hline
1 : Length & 0.300\\
2 : Radius &0.230\\
3 : Density & 0.227\\
4 : Begin & 0.103\\
5 : NInteractinglayer$/$Nlayer  & 0.080\\
6 : TrackMultiplicity & 0.060\\
\hline
\end{tabular}
\end{table}

\begin{table}[tbp]
\caption{\label{tab:rank_pie_datatraining} Variable ranking of separation importance in the case of~BDT$_{\pi e}$.}
\centering
\begin{tabular}{l||l}
\hline
\multirow{2}{*}{Rank : Variable} &\multirow{2}{*}{Variable relative weight}\\
 & \\
 \hline
1 : Radius &0.195\\
2 : NInteractinglayer$/$Nlayer  & 0.191\\
3 : Density & 0.189\\
4 : Length & 0.151\\
5 : Begin & 0.141\\
6 : TrackMultiplicity & 0.131\\
\hline
\end{tabular}
\end{table}

\section{Results}
\label{Para:results}
\begin{figure}[tbp]
\begin{center}
\begin{tabular}{ccc}
\includegraphics[width=0.49\textwidth]{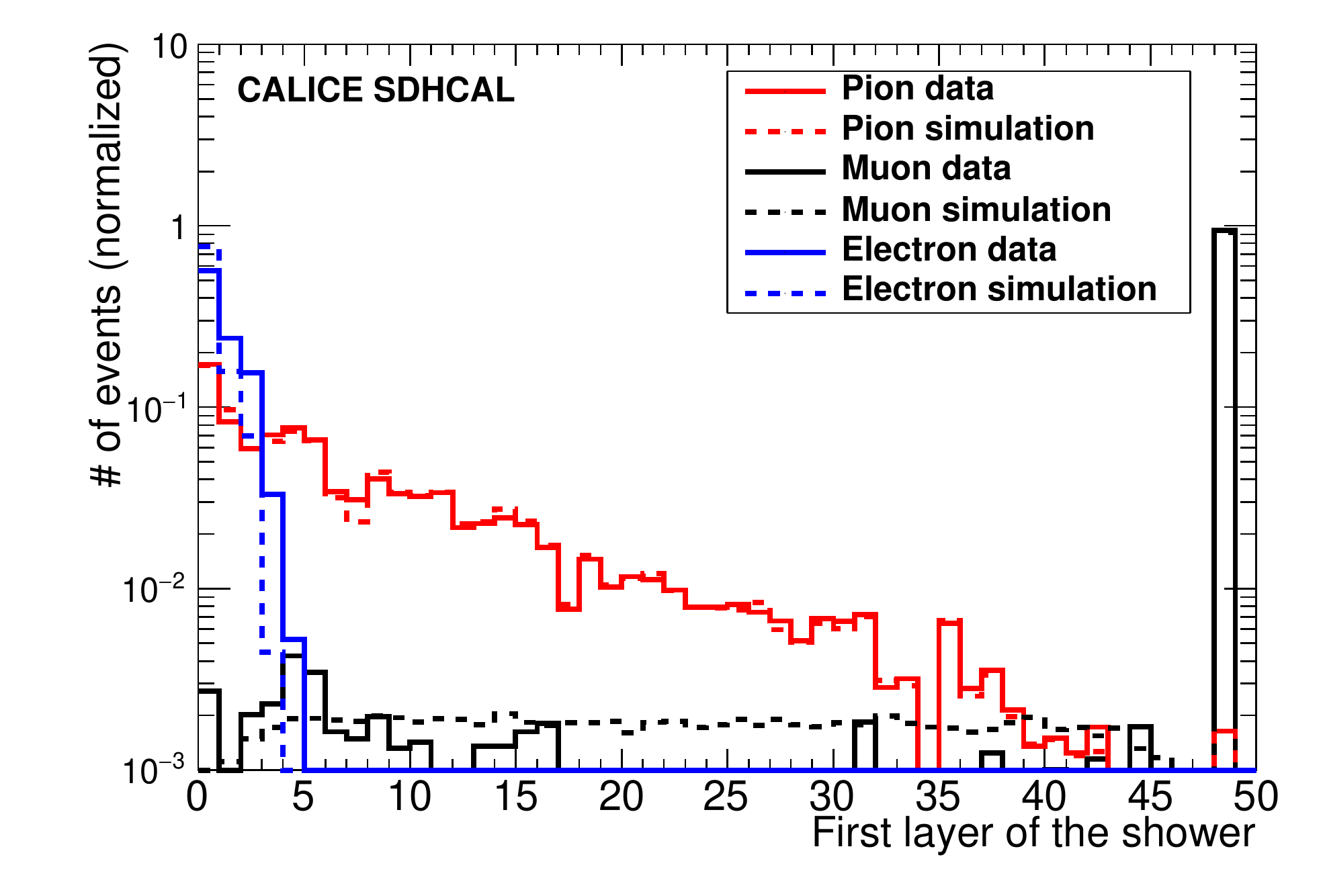}&
\includegraphics[width=0.49\textwidth]{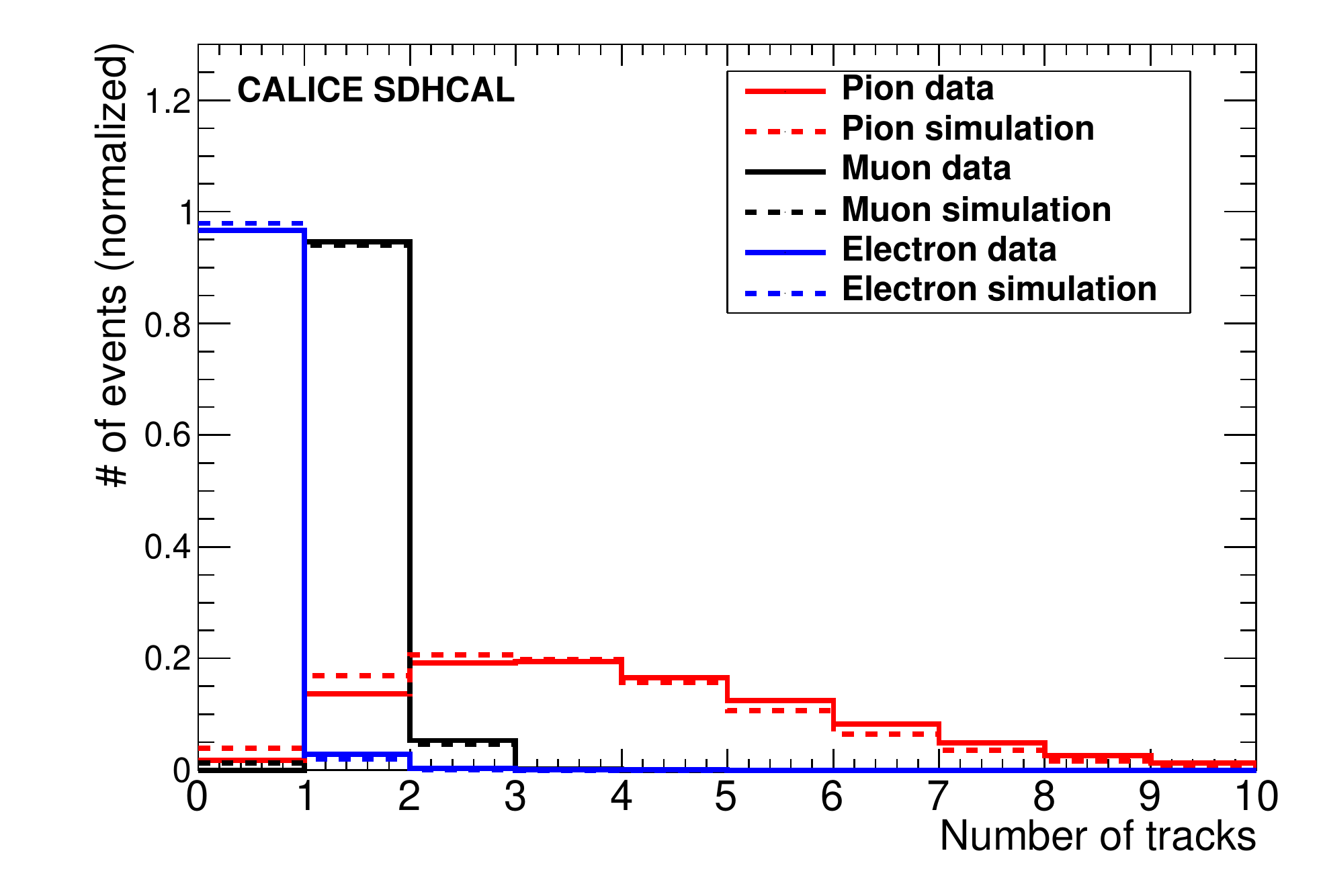}\\
\includegraphics[width=0.49\textwidth]{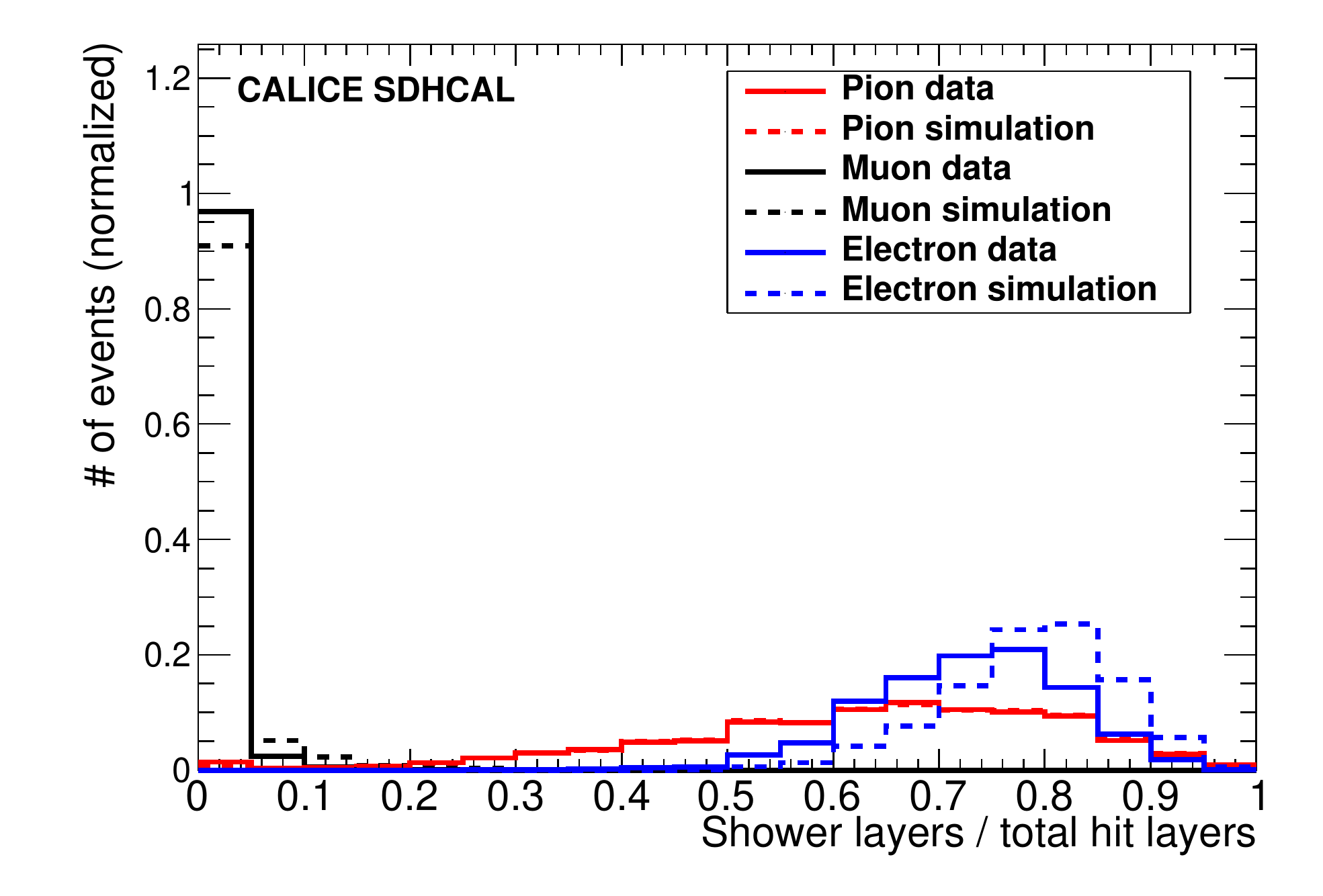}&
\includegraphics[width=0.49\textwidth]{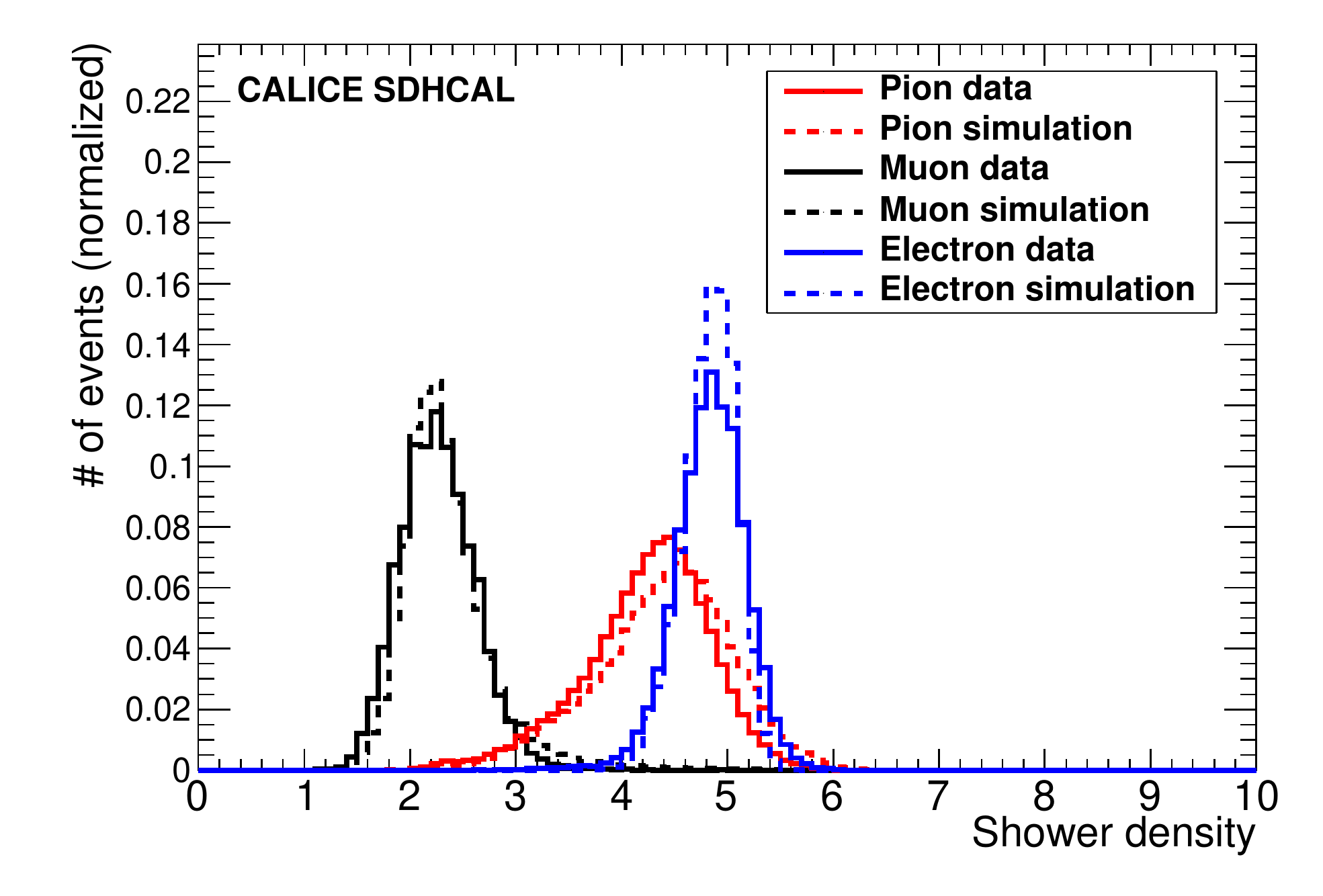}\\
\includegraphics[width=0.49\textwidth]{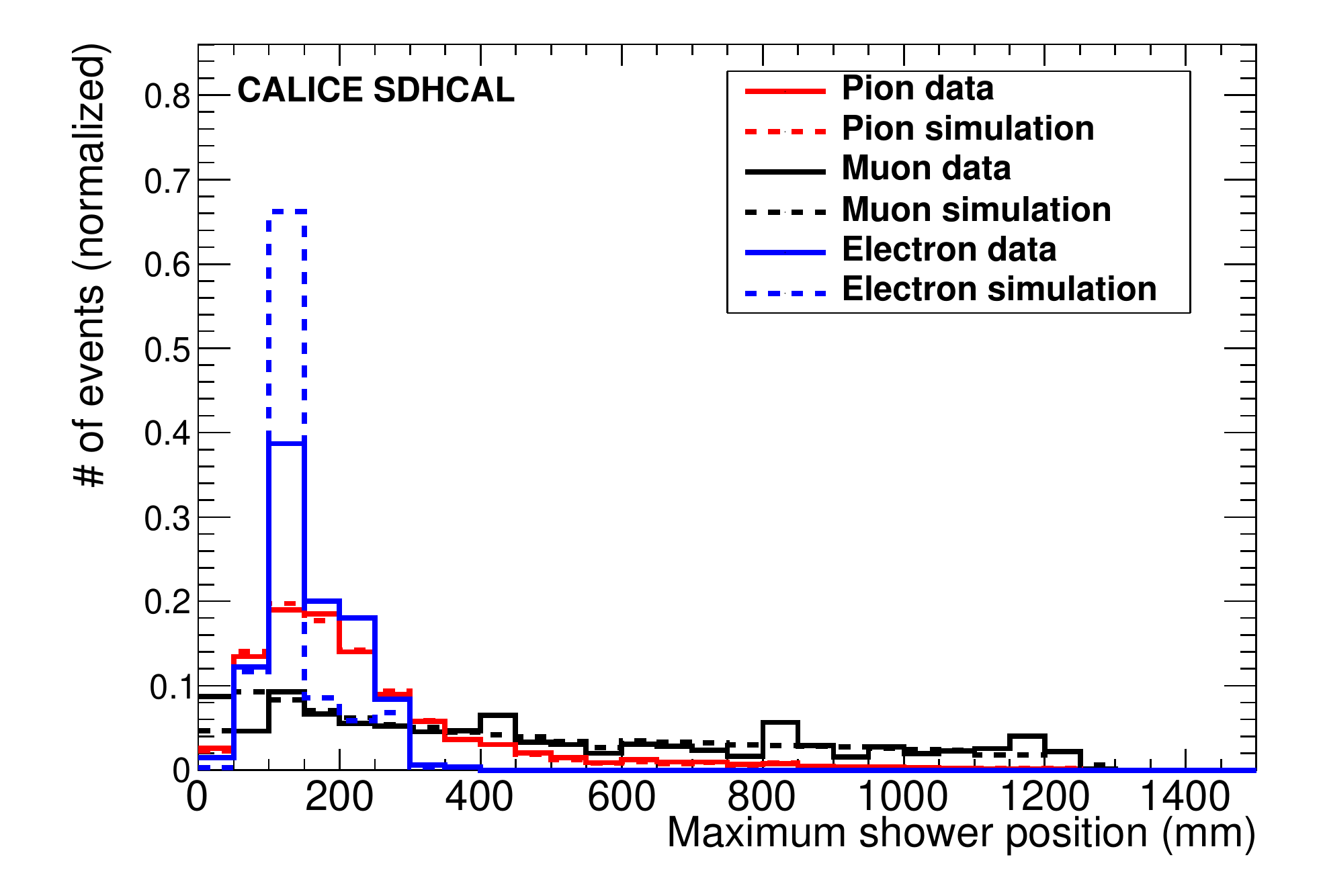}&
\includegraphics[width=0.49\textwidth]{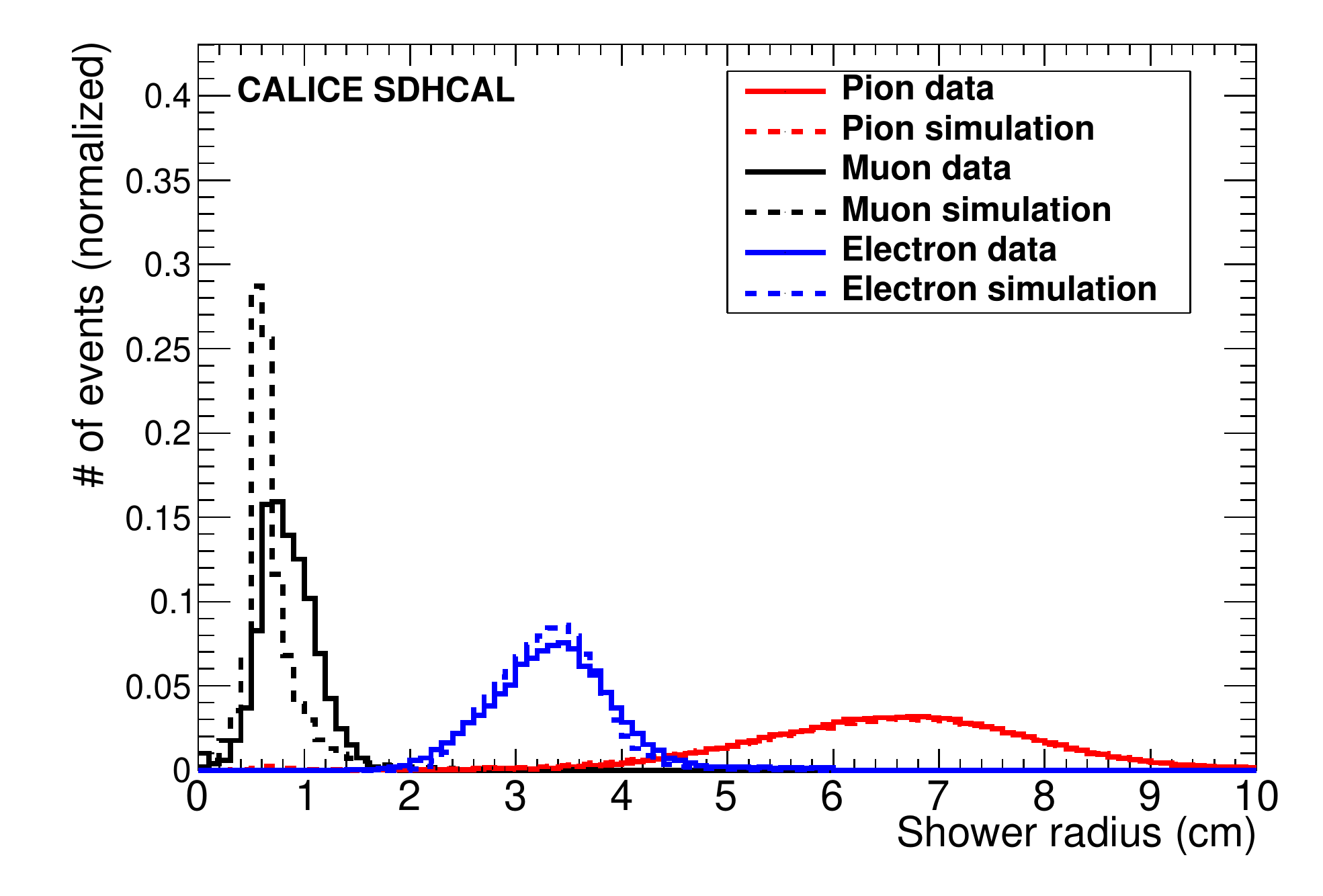}\\
\end{tabular}
\caption{Distributions of six input variables of electron, muon and pion samples. Continuous lines refer to data and dashed ones to the simulation. The pion samples are classified with the data-based training BDT method and others is obtained without applying BDT-based classifiers.}
\label{fig:Variables}
\end{center}
\end{figure}

\begin{figure}[tbp]
\begin{center}
\begin{tabular}{cc}
\includegraphics[width=0.49\textwidth]{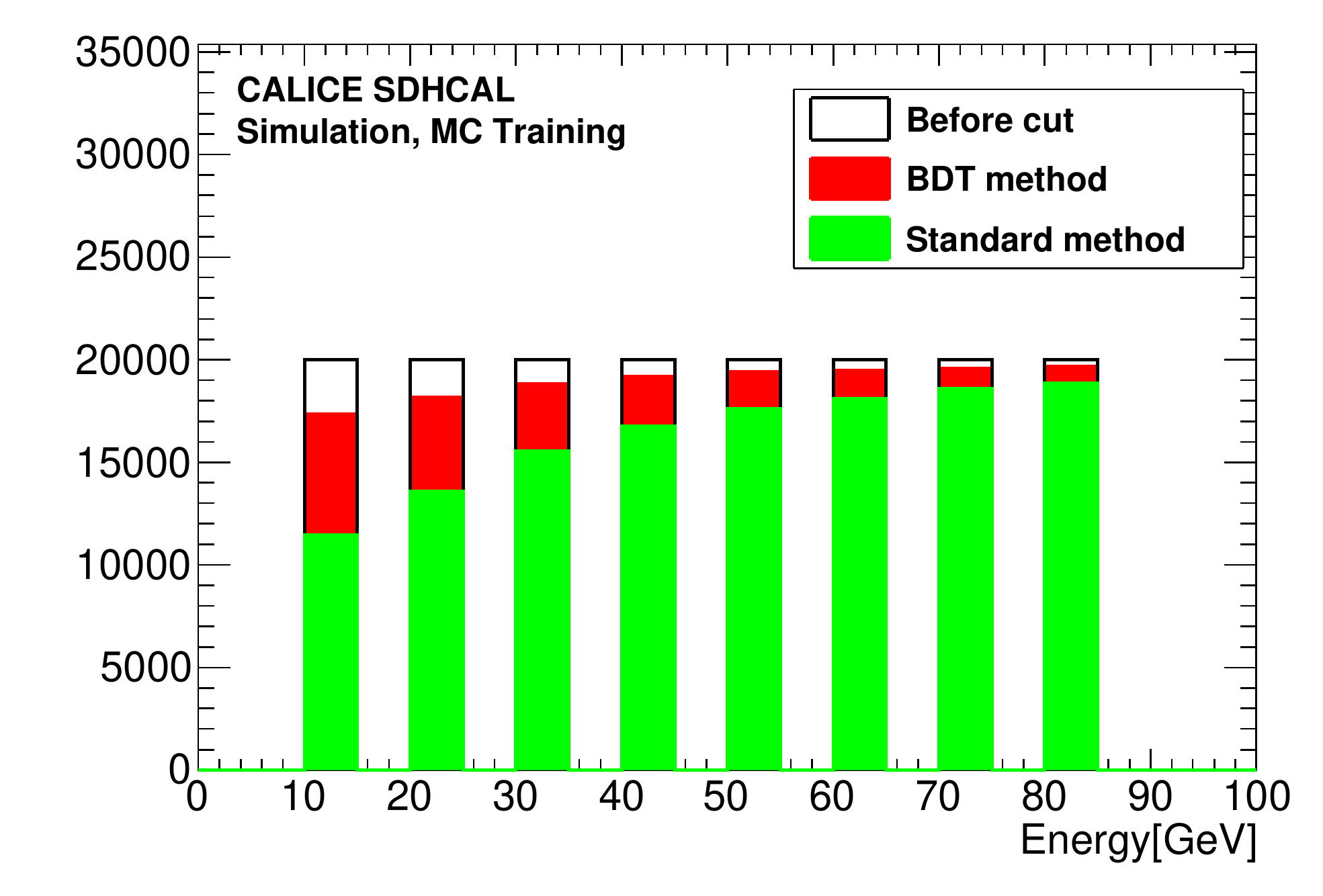}&
\includegraphics[width=0.49\textwidth]{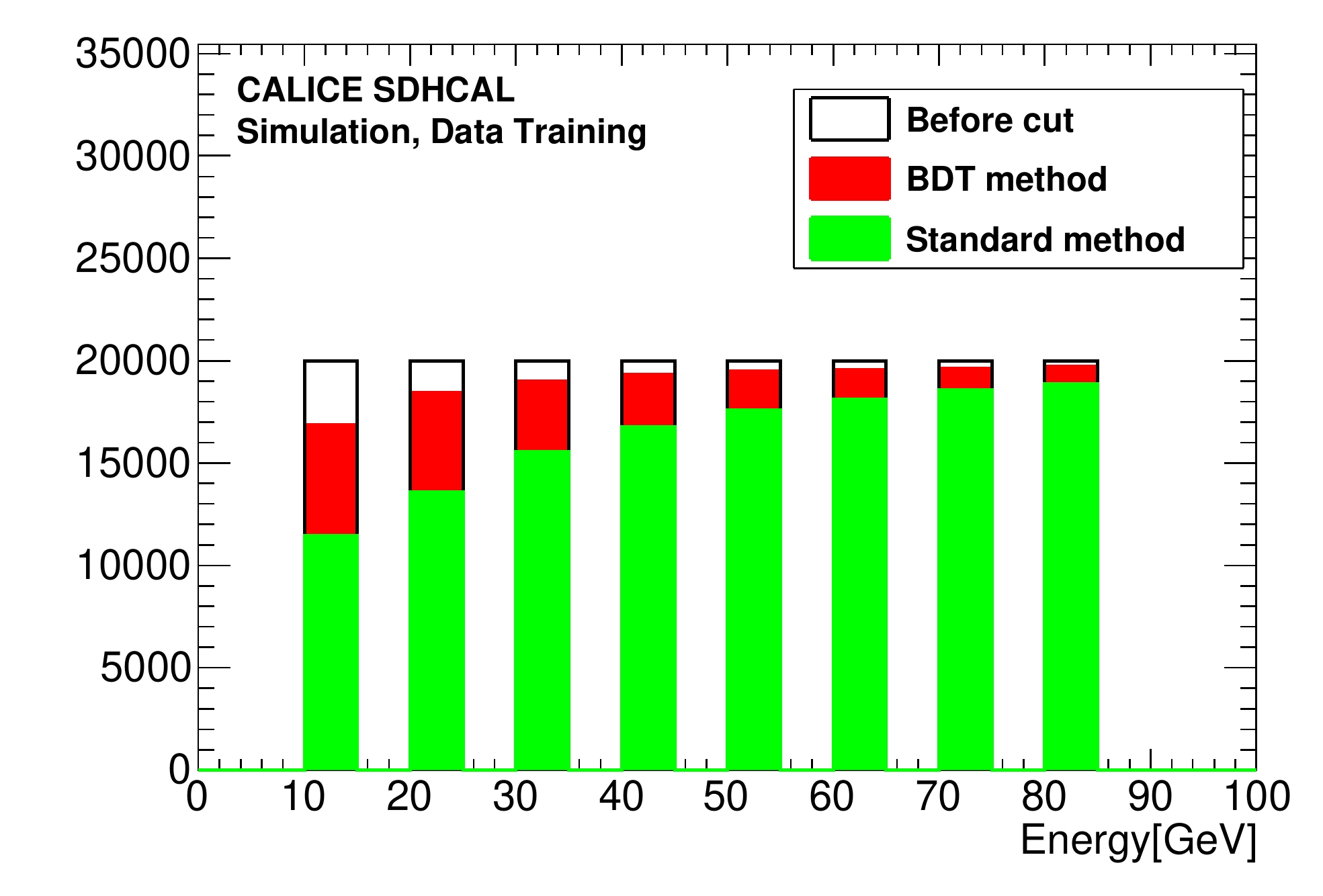}\\
\end{tabular}
\caption{The number of simulated events of different energy points from 10~GeV to 80~GeV before~(white) and after applying the standard method~(green) or BDT method~(red). The left plot shows the results from BDT method with MC Training approach while the right one shows the results with Data Training approach. }
\label{fig:number_MCTr}
\end{center}
\end{figure}

\begin{figure}[tbp]
\begin{center}
\begin{tabular}{cc}
\includegraphics[width=0.49\textwidth]{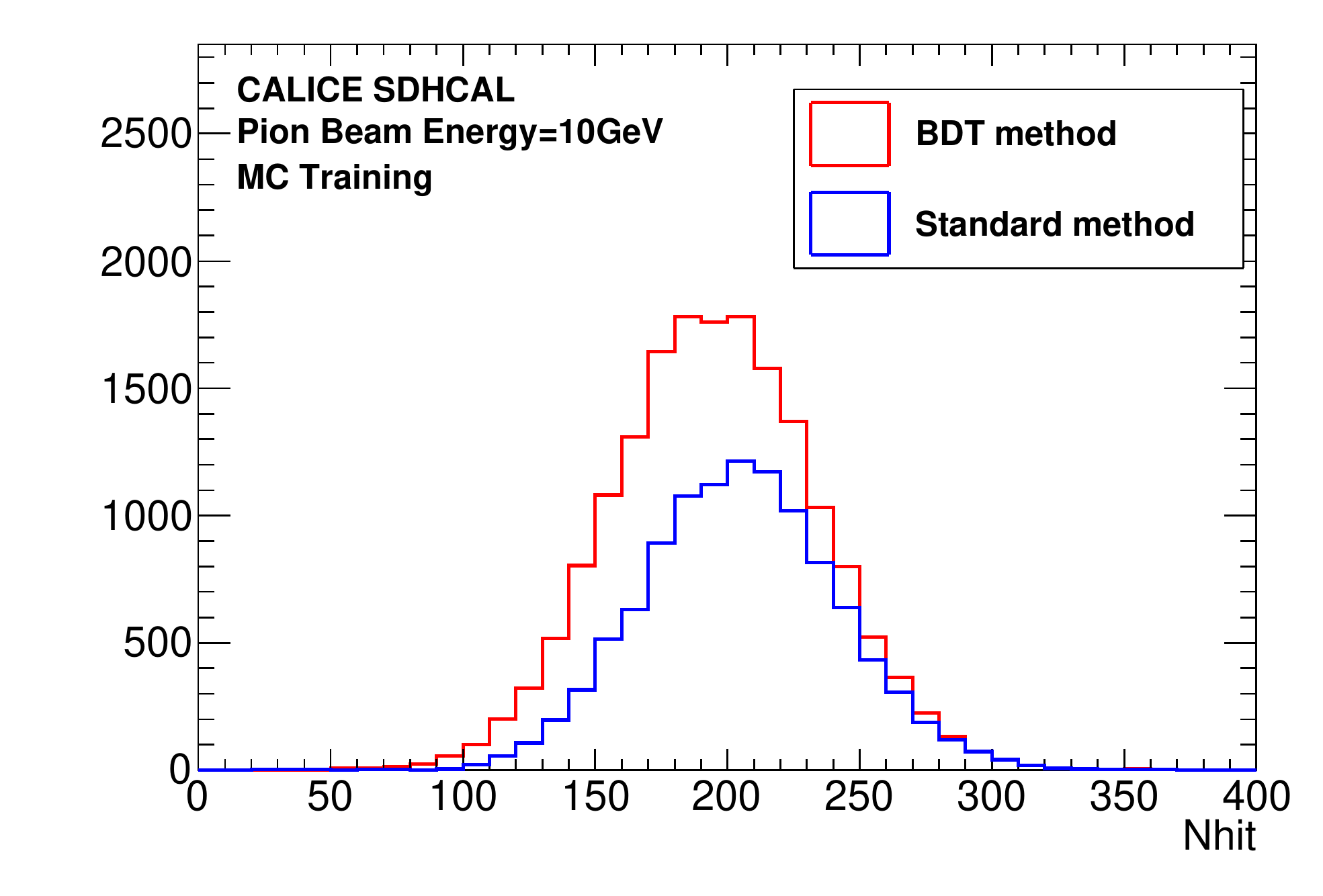}&
\includegraphics[width=0.49\textwidth]{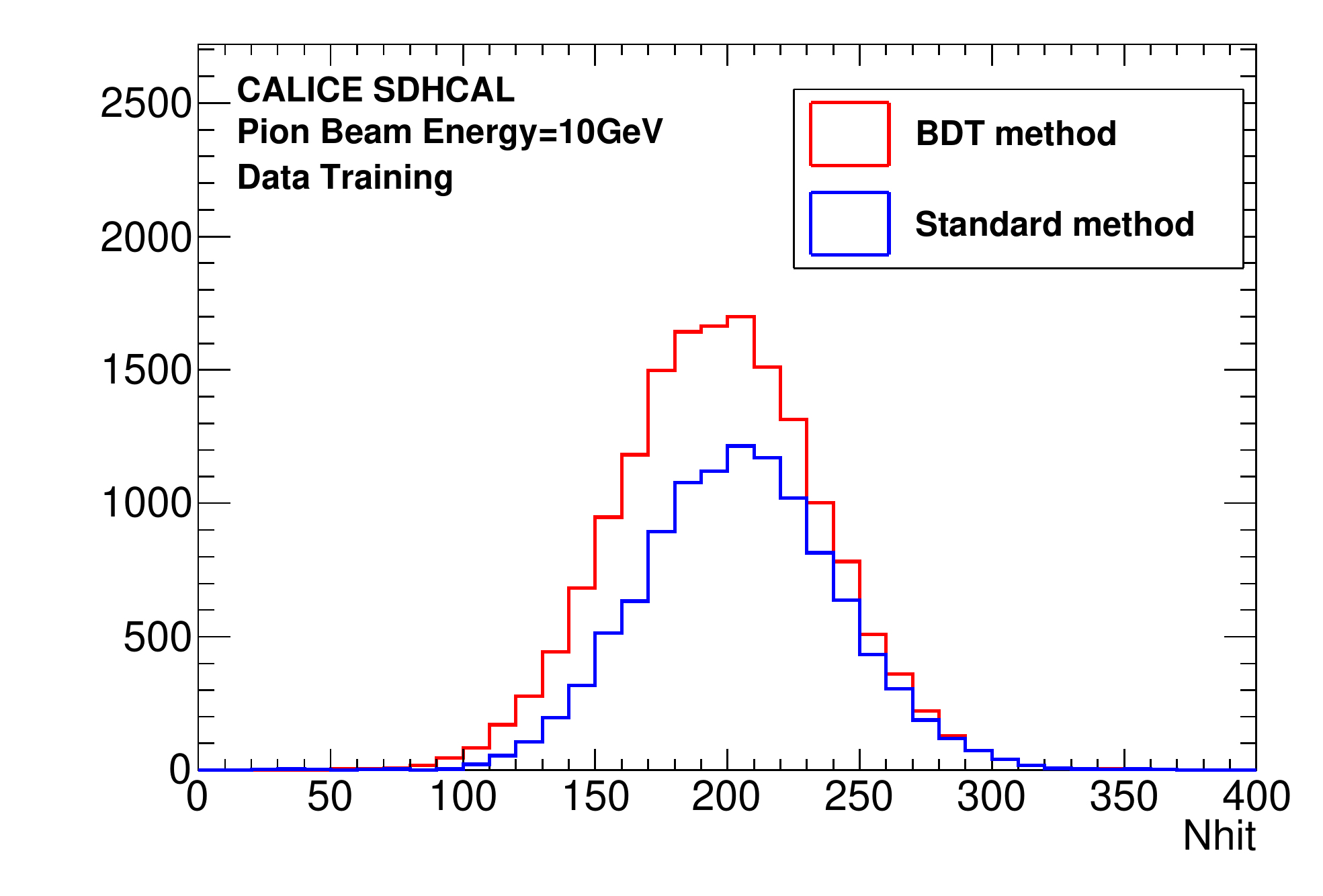}\\
\end{tabular}
\caption{ Distribution of the total number of hits for the 10~GeV pion beam data selected by the standard method~(blue) and the BDT method~(red). The left plot shows the results from BDT method with MC Training approach while the right one shows the results with Data Training approach. }
\label{fig:number_dataTr}
\end{center}
\end{figure}

The distributions of input variables for the data and simulation events of pion, muon and electron are shown in Fig.~\ref{fig:Variables}. Only the  pion data sample distributions are obtained after applying the data-based BDT classifiers. A good agreement between the data and simulation events for  pions is observed. It also confirms the power of the BDT method.
The rejection of muons and electrons presented in the pion data sample using the BDT allows us to have more statistics and a rather pure pion sample as explained in the previous section. Figure~\ref{fig:number_MCTr} shows the results of comparison in event selection between the standard method  and the BDT-based method  using the simulation samples.  For both simulation and beam data, the BDT method  leads to more statistics comparing to the standard method~\cite{FirstResults} in particular at low energy as shown in  Fig.~\ref{fig:number_dataTr} for the comparison of the selected events as a function of the total number of hits for the 10~GeV pion beam data. We also do not observe any significant deviation of energy resolution when applying the standard energy reconstruction described in Ref.~\cite{FirstResults} on the pion events selected by the BDT method.

\section{Conclusion}
\label{Para:conclusion}
A new particle identification method using BDT-based MVA technique is applied to purify the pion events collected at the SPS H2 beamline in 2015 by the CALICE SDHCAL prototype. The new method uses the topological shape of events associated to muons, electrons and pions in the CALICE SDHCAL to reject the two first species. A significant statistical gain  is obtained with respect to the standard method used in the work presented in Ref~\cite{FirstResults}. This statistical gain is particularly significant at energies up to 40~GeV and can be explained by the fact that the showers that start in the first layers are not all rejected. This gain shows the better efficiency and separation power of the multivariate approach over the cut-based approach  of the standard method.  The BDT-based particle identification in CALICE SDHCAL is a robust and a reliable method as confirmed by the results of two different training approaches. 
\section{Acknowledgements}
This study was supported by National Key Programme for S\&T Research and Development~(Grant NO.\ \: 2016YFA0400400).

\end{document}